\documentclass[11pt]{article}
\pdfoutput=1
\usepackage{graphicx}
\usepackage{hyperref}
\usepackage{epsfig}
\usepackage{amsmath}
\usepackage{amsthm}
\usepackage{amssymb}
\usepackage{multirow}
\usepackage{color}
\usepackage[nosort]{cite}
\usepackage{hyperref}
\usepackage[fontsize=11.45pt]{scrextend}
\usepackage{braket,mathrsfs,slashed}
\usepackage{float}
\usepackage{setspace}
\usepackage[caption=false]{subfig}
\newlength{\abstractwidth}
\newcommand{\be}{\begin{equation}}
\newcommand{\ee}{\end{equation}}

\renewcommand{\title}[1]{\vbox{\center\bf{\Large{#1}}}\vspace{5mm}}
\renewcommand{\author}[1]{\vbox{\center#1}\vspace{5mm}}
\newcommand{\address}[1]{\vbox{\center\em#1}}

\usepackage{dsfont}

\usepackage[utf8]{inputenc}
\usepackage{rotating}
\usepackage{dsfont}
\usepackage[paper=a4paper,margin=1.5cm]{geometry}
\usepackage[utf8]{inputenc}
\usepackage{rotating}
\usepackage{soul}

\usepackage{mathrsfs} 
\usepackage{bbm}
\usepackage{etoolbox} 
\usepackage{multirow}

\renewcommand\[{\begin{equation}}
\renewcommand\]{\end{equation}}

\newcommand{\ba}{\begin{eqnarray}}
\newcommand{\ea}{\end{eqnarray}}

\hypersetup{pdfstartview=FitV,colorlinks=true,linkcolor=midblue,citecolor=midblue,filecolor=midblue,urlcolor=midblue}

\definecolor{midblue}{rgb}{0,0,0.5}

\begin{document}
	
		\newgeometry{top=3.1cm,bottom=3.1cm,right=2.4cm,left=2.4cm}
		
	\begin{titlepage}
	\begin{center}
		\hfill \\
		\vskip 0.5cm

		\title{Scalar weak gravity bound from full unitarity}
		
			\author{\large Anna Tokareva$^{a,b,e\,\dagger}$, Yongjun Xu$^{a,c,d\,\ddagger}$ }
			
			\address{$^a$School of Fundamental Physics and Mathematical Sciences, \\Hangzhou Institute for Advanced Study, UCAS, Hangzhou 310024, China\\[1.5mm]
                $^b$International Centre for Theoretical Physics Asia-Pacific, Beijing/Hangzhou, China\\[1.5mm]
                $^c$Institute of Theoretical Physics, Chinese Academy of Sciences, Beijing 100190, China\\[1.5mm]
                $^d$University of Chinese Academy of Sciences, Beijing 100049, China\\[1.5mm]
                $^e$Department of Physics, Blackett Laboratory, Imperial College London, SW7 2AZ London, UK\\
                }
				\vspace{.3cm}

		\end{center}

\vspace{0.15cm}

\begin{abstract}
The weak gravity conjecture can be formulated as a statement that gravity must be the weakest force, compared to the other interactions in low-energy effective field theory (EFT). Several arguments in favor of this statement were presented from the side of string theory and black hole physics. However, it is still an open question whether the statement of weak gravity can be proven based on more general assumptions of causality, unitarity, and locality of the fundamental theory. These consistency requirements imply the dispersion relations for the scattering amplitudes, which allow us to bound the EFT coefficients. The main difficulty for obtaining these constraints in the presence of gravity is related to the graviton pole, which makes the required dispersion relations divergent in the forward limit. In this work, we present a new way of deriving the bound on the ratio between the EFT cutoff scale and Planck mass from confronting the IR divergences from the graviton pole and one-loop running of the EFT Wilson coefficient in front of the dimension-12 operator. Our method also allows the incorporation of full unitarity of the partial wave expansion of the UV theory. We examine the EFT of a single shift-symmetric scalar in four dimensions and find that the relation between the EFT coupling in front of the dimension-12 operator and the Planck mass.
\end{abstract}
\vspace{6cm}
\noindent\rule{6.5cm}{0.4pt}\\
$\,^\dagger$ \href{mailto:tokareva@ucas.ac.cn}{tokareva@ucas.ac.cn}\\	
$\,^\ddagger$ \href{mailto:xuyongjun23@mails.ucas.ac.cn}{xuyongjun23@mails.ucas.ac.cn}

\end{titlepage}

{
	\hypersetup{linkcolor=black}
	\tableofcontents
}

\baselineskip=17.63pt



\newpage

\section{Introduction}

Weak gravity conjecture is a criterion for effective field theory (EFT), which determines whether it can be consistently coupled to gravity. Its original formulation was given in the setup of $U(1)$ gauge theory coupled to gravity \cite{Arkani-Hamed:2006emk} as a condition on the charge-to-mass ratio in the Planck units,
\begin{equation}
  \frac{1}{M_P}<\frac{\sqrt{2} q}{m}.
\end{equation}
For the charged black holes, it means that every black hole must be able to evaporate without forming naked singularities caused by violating an extremality bound. This condition can also be seen as a requirement that gravity is the weakest force, as it actually bounds the gravitational coupling $1/M_P$ from above. In the framework of EFT of photon states, where all massive particles are integrated out, the photon couplings are proportional to the powers of charge-to-mass ratios. The weak gravity bound, thus, would relate gravitational coupling $1/M_P$ and Wilson coefficients in the EFT of photons. Significant efforts were made towards derivation of weak gravity conjecture from string theory \cite{Arkani-Hamed:2006emk,2401.14449,1903.06239,2201.08380,2409.10003,Bastian:2020egp}, black hole physics \cite{Cheung:2018cwt,Cottrell:2016bty,Hebecker:2017uix,Abe:2023anf,DeLuca:2022tkm,Barbosa:2025uau,Cao:2022ajt,Cao:2022iqh,Aalsma:2019ryi}, and from properties of the scattering amplitudes \cite{Bellazzini:2019xts,Hamada:2018dde,Arkani-Hamed:2021ajd,Henriksson:2021ymi,Alberte:2020bdz}. More arguments based on
holography~\cite{Nakayama:2015hga,Harlow:2015lma,Benjamin:2016fhe,Montero:2016tif}, cosmic censorship~\cite{Cottrell:2016bty,Horowitz:2016ezu,Crisford:2017gsb,Yu:2018eqq}, dimensional reduction~\cite{Brown:2015iha,Brown:2015lia,Heidenreich:2015nta,Heidenreich:2016aqi,Lee:2018urn}, and infrared consistency~\cite{Cheung:2014ega,Andriolo:2018lvp,Bittar:2024xuc}
were also provided in favor of the weak gravity conjecture.

In this work, we consider an EFT of shift-symmetric scalar coupled to gravity and address the question of whether we can derive a similar type of upper bound on the gravitational coupling set by the scalar EFT couplings. We assume that EFT of the shift-symmetric scalar field has a completion at high energies that satisfies the general properties of a consistent quantum theory, such as unitarity, causality, and locality. In a number of recent papers, the technique of deriving constraints on the EFT parameters from the desired properties of the UV theory has been developed \cite{Bellazzini:2019xts,Arkani-Hamed:2021ajd,Henriksson:2021ymi,Alberte:2020bdz,Saraswat:2016eaz,Guerrieri:2021ivu,Pham:1985cr, Pennington:1994kc,Nicolis:2009qm, Komargodski:2011vj,Remmen:2019cyz,Herrero-Valea:2019hde,Bellazzini:2017fep,deRham:2017avq,deRham:2017zjm,deRham:2017imi,Wang:2020jxr, Tokuda:2020mlf,Li:2021lpe,Caron-Huot:2021rmr,Du:2021byy,Bern:2021ppb,Li:2022rag, Caron-Huot:2022ugt,Herrero-Valea:2020wxz,EliasMiro:2022xaa,Bellazzini:2021oaj, Sinha:2020win, Trott:2020ebl,Herrero-Valea:2022lfd,Hong:2023zgm,Chiang:2022jep,Huang:2020nqy,Noumi:2021uuv, Xu:2023lpq, Chen:2023bhu,Noumi:2022wwf,deRham:2022hpx, Hong:2024fbl,Bern:2022yes,Ma:2023vgc,DeAngelis:2023bmd,Acanfora:2023axz,Aoki:2023khq,Xu:2024iao,EliasMiro:2023fqi,McPeak:2023wmq,Riembau:2022yse,Caron-Huot:2024tsk,Caron-Huot:2024lbf,Wan:2024eto,Buoninfante:2024ibt,Berman:2024owc,deRham:2025vaq,Berman:2025owb}. Typically, bounds on the EFT coefficients arise from dispersion relations for the scattering amplitudes which allow to relate them to positive-definite integrals and sums over partial waves. Their positive-definiteness is a consequence of the positivity of the imaginary parts of partial wave amplitudes of the UV theory. A set of integral inequalities that can be formulated for different combinations of EFT coefficients leads to compact bounds on the EFT parameter space. Although loop corrections and poles in the amplitude corresponding to the graviton exchange lead to the IR singularities, numerical and analytical methods of obtaining bounds in these more complicated situations were also developed \cite{Bellazzini:2021oaj,Riembau:2022yse,Bellazzini:2020cot,Beadle:2024hqg,Ye:2024rzr,Bertucci:2024qzt,Chang:2025cxc, Beadle:2025cdx, Pasiecznik:2025eqc,Ye:2025zhs}. If only positivity condition on partial waves is assumed, the weak gravity bound in the scalar case cannot be obtained  \cite{Andriolo:2020lul}, as well as weak gravity conjecture in EFT of photons coupled to gravity \cite{Henriksson:2022oeu}. It seems that more information about the UV theory is required. 

In this paper, we use the full unitarity condition on the partial wave amplitudes in four dimensions. This means that we assume the EFT is completed by a unitary four-dimensional theory. Under this condition, supplied by the convergence requirements for certain sums and integrals of partial waves, we find the bound relating the scalar EFT self-coupling to the Planck mass. We use Hölder’s inequality, which allows us to implement the full unitarity condition and construct the upper bound on gravitational coupling, making an explicit statement on how weak the gravity should be, compared to the scalar self-coupling. 

The paper is organized as follows. 
\begin{description}

\item[Sec.~\ref{sec:dispersion}:] We formulate dispersion relations in the presence of graviton pole and express the EFT couplings through the moments of positive-definite sums and integrals over partial waves.

\item[Sec.~\ref{sec:hodler}:] We describe the construction of Hölder’s inequality, which allows to obtain the weak gravity bound in the EFT of shift-symmetric scalar, and present a particular realization of the method.

\item[Sec.~\ref{sec:smeared}:] We introduce the smeared dispersion relations leading to positive-definite functionals, and show how the stronger bound on the gravitational coupling can be obtained in the limit when the smearing region goes to zero.

\item [Sec.~\ref{sec:conclusions}:]  
We discuss the validity of our assumptions about the UV theory and possible implications of our results, as well as directions for future study.
\end{description}

\section{Dispersion relations with the graviton pole and EFT loops}
\label{sec:dispersion}
An issue of divergence of the graviton exchange amplitude in the forward limit makes it more difficult to formulate a technique of dispersion relations between UV and IR limits of the complete theory. Many different approaches to the problem were formulated before, starting from the assumption that the graviton pole divergence can be ignored in dispersion relations (this could sound reasonable if all scales in EFT are much smaller than Planck mass), which led to certain conclusions about QED and Standard Model \cite{Alberte:2020bdz,Aoki:2021ckh}. However, more accurate treatment of this IR divergence states that it should be reproduced in dispersion relations from a specific behavior of the amplitude in the UV theory, thus, imposing certain constraints on the UV, such as Regge trajectory \cite{Tokuda:2020mlf,Herrero-Valea:2022lfd,Noumi:2021uuv,Noumi:2022wwf,Alberte:2021dnj,deRham:2022gfe} and even stronger conditions imposed by unitarity which are shown to be satisfied for the eikonal model of graviton-mediated scattering \cite{Haring:2022cyf,Haring:2024wyz}. 

\begin{figure}[H]  
\label{fig:contour}
    \centering
\includegraphics[width=0.5\textwidth]{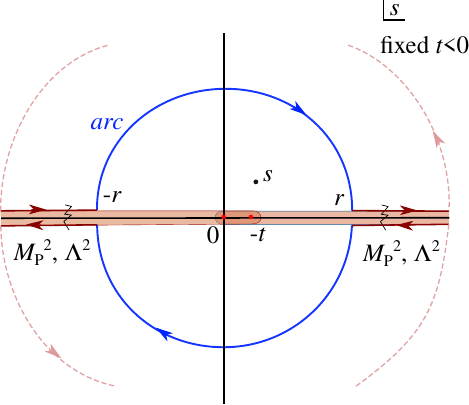}  
    \caption{ Integration contour used in \eqref{arc} and branch cut singularities of the scattering amplitude at fixed real $t<0$. Here $\Lambda$ is an EFT cutoff scale, and $r<\Lambda$ is the radius of the arc.}  
    \label{fig: 1}  
\end{figure}

We make use of the dispersion relation based on the analyticity and locality of the full scattering amplitude including the graviton exchange. For fixed $-t>0$ we can relate an integral over the relatively small arc (which can be computed in the EFT) to the integrals over branch cuts of the amplitude, see Figure \ref{fig:contour}. 

\begin{equation}
\label{arc}
   \frac{1}{2\pi i} \oint_{arc}\frac{A(\mu,t)}{(\mu-s)^{n+1}}=\int_{r }^{\infty} \frac{d \mu}{\pi} \frac{\operatorname{Disc}_s A(\mu, t)}{(\mu-s)^{n+1}}+ (-1)^{n} \int_{r -t}^{\infty} \frac{d \mu}{\pi} \frac{\operatorname{Disc}_u A(\mu, t)}{(\mu-u)^{n+1}}+\frac{1}{2\pi i} \oint_{\infty}\frac{A(\mu,t)}{(\mu-s)^{n+1}}.
\end{equation}

If we assume the Schwarz reflection principle, $A(\mu,t)=A^*(\mu^*,t)$, we have,
\begin{equation}
    \operatorname{Disc}_s A(\mu,t) = \text{Im} A(\mu,t).
\end{equation}
The last term in \eqref{arc} is a boundary term, which vanishes if the amplitude is bounded $A_{\text{mp}} < s^2$ in the limit of $ |s| \to \infty $ at fixed $ t$. This condition is a consequence of unitarity in theories with a mass gap, as it follows from the Martin-Froissart bound \cite{Froissart:1961ux, Martin:1962rt, Jin:1964zz}. If graviton exchange is taken into account, this condition has to be assumed additionally; it is not possible to derive it from the first principles. However, under several reasonable assumptions, it can be obtained as a result of eikonal resummation \cite{Haring:2022cyf}.

We define a set of quantities that can be computed in the EFT theory,
\begin{equation}
   \label{M}
   M_n(t)\equiv \frac{1}{2\pi i} \oint_{arc}\frac{A(\mu,t)}{\mu^{n+1}}+(-1)^{n} \int_{r }^{r -t} \frac{d \mu}{\pi} \frac{\operatorname{Disc}_u A(\mu, t)}{(\mu-u)^{n+1}}.
    \end{equation}
The dispersion relation \eqref{arc} allows us to rewrite them through integrals over discontinuities of the UV amplitude. These discontinuities can be expanded in partial waves,
\begin{equation}
\text{Disc}_s A(\mu,t) = 16\pi \sum_{j=\text{even}}^{\infty} (2j+1) \, \text{Im} f_j(\mu) P_j\left( 1 + \frac{2t}{\mu} \right).
\end{equation}

where \( P_j \) are the Legendre polynomials , and \( f_j \) are the partial-wave expansion coefficients. They must satisfy the full unitarity condition $0\leq {\rm Im}\,f_j \leq2$. This way, the left-hand side (determined by the EFT) of the dispersion relation appears to be the integrated sum of positive-definite and bounded functions. This way, constraints on the Wilson coefficients of EFTs emerge from a set of integral inequalities formulated for bounded and positive-definite functions \cite{Bellazzini:2020cot,Arkani-Hamed:2020blm}. A systematic account of all these constraints can be done based on the theory of moments \cite{Chiang:2021ziz,Chiang:2022jep,Chiang:2022ltp}. 

This well-developed technique needs modification in the presence of massless loops and graviton pole divergence \cite{Bellazzini:2021oaj,Beadle:2024hqg}. The reason for that is related to the fact that partial wave sums, as well as integrals, may not be convergent in certain cases. Thus, only selected combinations are well-defined mathematically. In dimensions higher than four, this problem can be avoided by introducing certain smearing functions that are still positive-definite \cite{Caron-Huot:2021rmr}. However, in four dimensions, there are no proper positive-definite functions, so overcoming this issue is more challenging. In this paper, we develop a method based on smearing functions, which allows us to define convergent moment integrals properly addressing their analytic properties in the limit $t\rightarrow 0$.


\subsection{Scattering amplitude in EFT}

In this paper, we consider the scattering of four massless scalars described by the EFT of the shift-symmetric scalar field,
\begin{equation}
\label{amp}
    A(s,t,u) = A_S (s,t,u)+ A_G(s,t,u).
\end{equation}
Here $s,~t,~u$ are Mandelstam variables and $u=-s-t$ on-shell. The scalar self-interaction part of the amplitude $A_S$ is given by,
\begin{equation}
\begin{split}
    A_S(s,t,u) = &\ g_2 \left( s^2 + t^2 + u^2 \right) + g_3 s t u + g_4 \left( s^2 + t^2 + u^2 \right)^2 \\
    &+ \left[ s^2 \left( b_2 t u + b_1 s^2 \right) \log (-s) + (s - t) + (s - u) \right] \\
    &+ g_5 (stu)(s^2+t^2+u^2) + \left[ s^3 \left( c_2 t u + c_1 s^2 \right) \log (-s) + (s - t) + (s - u) \right] +\cdots.
\end{split}
\end{equation}
Here we included the first loop corrections emerging from the scalar self-interaction. The corresponding $\beta$-functions are
\begin{equation}
    b_1=-\frac{21g_2^2}{240\pi^2}, \quad b_2=\frac{g_2^2}{240\pi^2}, \quad c_1=-\frac{g_2g_3}{60\pi^2}, \quad c_2=-\frac{g_2 g_3}{240\pi^2}.
    \end{equation}
The amplitude from a single graviton exchange $A_G$ can be written in a fully crossing-symmetric form as
\begin{equation}
    A_G(s,t,u) = \frac{1}{2 M_{\text{P}}^2} \left( \frac{t u}{s} + \frac{s t}{u} + \frac{s u}{t} \right).
\end{equation}
Based on the expression \eqref{amp}, we can compute the arc integrals $M_n$ defined in \eqref{M}. We provide here the expressions for them expanded at small $t$ up to the leading non-analyticities caused by the logarithmic corrections.

\begin{align}
&M_2= -\frac{1}{2 t M_{\text{P}}^2}
+ \frac{1}{3} \left(3 b_1 r^2 + 2 c_1 r^3 + 6 g_2 \right) \nonumber \\
&\quad + \frac{1}{2} t \left( -6 b_1 r - 4 b_2 r - 3 c_1 r^2 - 2 c_2 r^2 - 2 g_3 \right) \nonumber \\
&\quad + t^2 \left( 6 b_1 \log r + b_2 \log r - b_2 \log(-t) + 6 c_1 r + c_2 r + 12 g_4 \right) \nonumber \\
&\quad + \frac{t^3}{r} \left( 10 b_1 + 3 b_2 - 10 c_1 r \log r - c_2 r \log(-t) - 3 c_2 r \log r - 4 g_5 r \right) \nonumber \\
&\quad  + O(t^4). \label{eq:M2}
\end{align}



Here we notice that due to the presence of $\log(-t)$ terms, certain $t$-derivatives of $M_2$ may lead to $1/t$ divergencies. The graviton pole contributes only to $M_2$, however, this contribution makes it problematic to take $t\rightarrow 0$ limit in the dispersion relation, as well as find out the source of this divergence in the UV integrals over the discontinuities,
\begin{equation}
    \label{B}
   M_2(t)= \left[\left( \frac{1}{\mu^{3}}+ \frac{1}{(\mu+t)^{3}}\right)P_j\left(1+\frac{2t}{\mu}\right)\right], 
\end{equation}
   where we define the brackets as
       \begin{equation}
\label{brackets}
   [ X(\mu, j)]\equiv \sum_{j\,\rm{even}} 16(2j+1) \int_{r}^\infty {\rm d}\mu \,{\rm Im}\,f_j(\mu)X (\mu, j).   
\end{equation}
Given that Legendre polynomials are analytic in $t$ at $t\rightarrow 0$, the naive expansion of the brackets would not match with the IR theory structure. There is no contradiction because, for example, the partial wave expansion may not be convergent pointwise. Thus, the infinite sum can converge at any finite $t<0$, while being divergent at $t=0$. 

In this work, we follow the idea that the analytic structure of the IR theory imposes constraints on its UV completion. In particular, we compare the divergence \( \frac{1}{t} \) arising from gravitational exchange with the divergence that originates from the derivative with respect to the one-loop beta function coefficient \( b_2 \).

\subsection{Gravitational coupling through the moments of UV theory}

From the IR theory, we know that \( M_2 \) must exhibit a \( \frac{1}{t} \) divergence, regardless of the specific local UV completion. This divergence, arising from graviton exchange, is dominant compared to any quantity.
\begin{equation}
\label{pole}
    M_2=\frac{1}{2(-t) M_P^2},\quad t\rightarrow 0.
\end{equation}
the corresponding bracket is:
\begin{equation}
F\big|_{t \to 0} =\Big[ \frac{2}{ \mu^{3}} \Big] + \mathcal{O}(t),
\end{equation}
One can notice that there are other non-analyticities caused by massless loops from the scalar self-interactions. The main idea of this paper, roughly, is to relate these non-analyticities with the graviton pole divergence, in such a way that the divergencies cancel each other, although they are coming from different sectors of the theory. 

\subsection{Expressions for running of $g_4$ through moments of UV theory}

As we are looking for the relation between the beta function and the Planck mass, we also need to construct the expressions for them in terms of moments. To isolate the subleading logarithmic behavior, we begin by multiplying \( M_2 \) by \( t \) and then differentiating with respect to \( t \). This procedure removes the leading \( \frac{1}{t} \) divergence that arises from graviton exchange. To meaningfully compare the remaining \( \log(t) \) divergence with the original \( \frac{1}{t} \) term, at least four derivatives are required. In our analysis, we take eight derivatives with respect to \( t \); the reason for this choice will be explained in Section~\ref{sec:hodler}.

We can extract the leading coefficient as
\begin{equation}
F_2 \equiv \frac{\partial^8}{\partial t^8} \left( t M_2(t) \right) 
= -\frac{144 b_2}{t^5} + \frac{144 c_2}{t^4} + \mathcal{O}\left(\frac{1}{t^3} \right), \quad t \to 0.
\end{equation}

From the UV theory, \( F \) can also be written in terms of moments as
\begin{equation}
F_2 = \frac{\partial^8}{\partial t^8} \left[ \left( \frac{t}{\mu^3} + \frac{t}{(\mu + t)^3} \right) P_j\left(1 + \frac{2t}{\mu} \right) \right],
\end{equation}
Expanding this expression in the limit \( t \to 0 \), we obtain
\begin{equation}
\begin{aligned}
F_2\big|_{t \to 0} =\; & \Big[ \frac{1}{630 \mu^{10}} \Big(
2 j^{14} + 14 j^{13} - 329 j^{12} - 2156 j^{11} + 24675 j^{10} + 143472 j^9 \\
&\quad - 997847 j^8 - 4876508 j^7 + 21363839 j^6 + 81786194 j^5 \\
&\quad - 221210724 j^4 - 584720136 j^3 + 912065184 j^2 \\
&\quad + 1218913920 j - 914457600 \Big) \Big] + \mathcal{O}(t) \\
&\equiv \Big[ \frac{\phi(j)}{\mu^{10}} \Big] + \mathcal{O}(t),
\end{aligned}
\end{equation}
For convenience, we define the function \( \phi(j) \) to capture the angular momentum dependence of the UV contribution. However, it is important to note that \( \phi(j) \) is not positive definite for all values of \( j \). This poses a potential challenge when applying Hölder's inequality, which requires positivity.
\begin{figure}[h]
    \centering
    \includegraphics[width=0.5\linewidth]{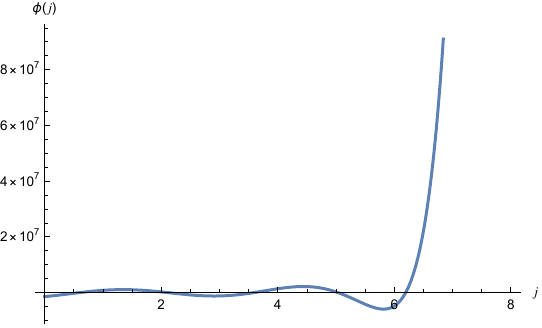}
    \caption{The value of $\phi(j)$ as a function of $j$; the picture shows that it is not positive definite for small $j$.}
    \label{fig:refined}
\end{figure}
Nevertheless, since we are ultimately interested in the behavior in the \( t \to 0 \) limit—corresponding to contributions from asymptotically large angular momentum — we can safely subtract a finite number of low-spin contributions from both sides of the inequality. This subtraction does not affect the validity of the inequality in the \( t \to 0 \) limit.

Specifically, we subtract the UV spectral contributions from angular momenta \( j = 0 \) up to some finite cutoff \( j_0 - 2 \). We then compare only the remaining parts of both sides, which dominate the behavior in the high-energy limit and ensure the applicability of Hölder’s inequality.
We will choose \( j_0 = 8 \) as a conservative cutoff. While larger values of \( j_0 \) can also be considered, taking \( j_0 \to \infty \) is not allowed, since in that limit the Legendre polynomial \( P_j\left(1 + \frac{2t}{\mu} \right) \) may no longer be positive definite. For our purposes, \( j_0 = 8 \) is a sufficiently safe and practical choice.

Under the full unitarity condition \( \text{Im} f_j(\mu) \le 2 \), the subtracted contributions are finite. We choose \( j_0 = 8 \). Then, for the second moment \( M_2 \), we have:
\begin{equation}
\begin{split}
M_2 \Big|_{0 \le j \le j_0 - 2} 
&= \sum_{\substack{j\,\text{even} \\ 0 \le j \le j_0 - 2}} 
16(2j+1) \int_{r}^\infty \mathrm{d}\mu \, \text{Im} f_j(\mu) \cdot \frac{2}{\mu^3} \\
&\leq \sum_{\substack{j\,\text{even} \\ 0 \le j \le j_0 - 2}} 
16(2j+1) \int_{r}^\infty \mathrm{d}\mu \, 2 \cdot \frac{2}{\mu^3} \\
&= \frac{896}{r^2}.
\end{split}
\end{equation}

Similarly, for the eighth derivative \( F_2 \), we find:
\begin{equation}
\begin{split}
F_2 \Big|_{0 \le j \le j_0 - 2} 
&= \sum_{\substack{j\,\text{even} \\ 0 \le j \le j_0 - 2}} 
16(2j+1) \int_{r}^\infty \mathrm{d}\mu \, \text{Im} f_j(\mu) \cdot \frac{\phi(j)}{\mu^{10}} \\
&\leq \sum_{\substack{j\,\text{even} \\ 0 \le j \le j_0 - 2}} 
16(2j+1) \int_{r}^\infty \mathrm{d}\mu \, 2 \cdot \frac{|\phi(j)|}{\mu^{10}} \\
&= \frac{276971520}{r^9}.
\end{split}
\end{equation}

\section{Hölder’s inequality in the forward limit}
\label{sec:hodler}
Now we are in a position to construct an inequality relating $[1/\mu^3]$ and $[\phi(j)/\mu^{10}]$ where $\phi(j)\propto j^{14}$ for large $j$ and is positive for all discrete $j$. We use here the full unitarity condition $0<{\rm Im}\,f_j<2$, and the generalized Hölder’s inequality, which states,

\begin{equation}
16 \sum_{j=j_0}^{\infty} \int_r^\infty f(\mu) g(\mu) \, d\mu \leq \left( 16 \sum_{j=j_0}^{\infty} \int_r^\infty f^q(\mu) \, d\mu \right)^{\frac{1}{q}} \left( 16 \sum_{j=j_0}^{\infty} \int_r^\infty g^p(\mu) \, d\mu \right)^{\frac{1}{p}},
\end{equation}
where \( p > 0 \), \( q > 0 \), and \( \frac{1}{p} + \frac{1}{q} = 1 \). This inequality allows us to relate integrals involving functions \( f(\mu) \) and \( g(\mu) \) to their respective powers \( f^q(\mu) \) and \( g^p(\mu) \).

\subsection{The method}

The main idea of application of Hölder’s inequality relating the two moments is to adjust the powers and functions such that we get a power of $[\phi(j)/\mu^{10}]$ as one of the integrals in the right hand side, while the other integral is just a number which doesn't contain the partial wave amplitudes. We propose to use the following positive functions \( f(\mu) \) and \( g(\mu) \),
\begin{equation}
\begin{split}
f(\mu) &= \left( \text{Im} f_j \right)^{\frac{1}{q}} (2j+1)^{\frac{1}{q}} \left( \frac{\phi(j)}{\mu^{10}} \right)^{\frac{1}{q}}, \\
f^q(\mu) &= \left( \text{Im} f_j \right) (2j+1) \left( \frac{\phi(j)}{\mu^{}} \right), \\
g(\mu) &= (2j+1)^{\frac{1}{p}} \frac{1}{\mu^{3 - 10 \frac{1}{q}}} \phi(j)^{-\frac{1}{q}}, \\
g^p(\mu) &= (2j+1) \frac{1}{\mu^{p(3 - 10 \frac{1}{q})}} \phi(j)^{-\frac{p}{q}}.
\end{split}
\end{equation}

Using these definitions, we can express the left-hand side of the Hölder’s inequality,

\begin{equation}
\frac{a}{2 M_{\text{P}}^2(-t)} = \left[ \frac{a}{\mu^3} \right] = 16\pi \sum_{j=j_0}^{\infty} \int_r^\infty (a \, \text{Im} f_j)(2j+1) \frac{1}{\mu^3} \, d\mu \leq 16\pi \sum_{j=j_0}^{\infty} \int_r^\infty \left( \text{Im} \,f_j \right)^{\frac{1}{q}} (2j+1) \frac{1}{\mu^3} \, d\mu.
\end{equation}

Here, we apply the inequality only in the limit of very small negative \( t \), and thus neglect the contributions from angular momenta \( j = 0, 2, \ldots, j_0 - 2 \). The parameter \( a \in \mathbb{R} \) is introduced to optimize the bound and is subject to the condition
\[
a \, \text{Im} f_j < \left( \text{Im} f_j \right)^{\frac{1}{q}} \quad \text{for all} \quad 0 \leq \text{Im} f_j \leq 2.
\]

This implies that \( a < \left( \text{Im} f_j \right)^{-\frac{1}{p}} \). Since the function \( \left( \text{Im} f_j \right)^{-\frac{1}{p}} \) is monotonically decreasing over the interval \( [0, 2] \), the parameter \( a \) must obey the upper bound

\[
0 \leq a \leq 2^{-\frac{1}{p}}.
\]

We retain the parameter \( a \) explicitly, as this allows for an additional optimization of the final constraint with respect to its value.

The right-hand side of the inequality becomes,
\begin{equation}
\left( 16 \sum_{j=j_0}^{\infty} \int_r^\infty \left( \text{Im} \,f_j \right) (2j+1) \frac{\phi(j)}{\mu^{10}} \, d\mu \right)^{\frac{1}{q}} \left( 16\sum_{j=j_0}^{\infty} \int_r^\infty (2j+1) \frac{1}{\mu^{p(3 - 10 \frac{1}{q})}} \phi(j)^{-\frac{p}{q}} \, d\mu \right)^{\frac{1}{p}}.
\end{equation}

The first multiplier is exactly the expression for $b_2$, while the last is just a number which can be computed, provided $\phi(j)$, $p$ and $q$. However, both the sum and integral should converge. This imposes the condition \( p(3 - 10/q) > 1 \) for the integral, and  \( 1 - 14q/p <-1  \) for the sum over $j$. Remarkably, it is crucial to have a power of $j^{14}$ in $\phi(j)$ because the proposed method would not provide any bound for lower power of $j$ in the moment. These requirements provide the following constraints,
\begin{equation}
p < 5 \quad \text{and} \quad p < \frac{9}{7}, \quad \text{or equivalently,} \quad \frac{9}{2} < q.
\end{equation}

Thus, using the relation \( \frac{1}{p} + \frac{1}{q} = 1 \), we obtain the inequality,
\begin{equation}
\frac{a}{2 M_{\text{P}}^2(-t)} < \left(- \frac{b_2}{t^5} \right)^{\frac{1}{q}} \left( 16 \sum_{j=j_0}^{\infty} \int_r^\infty (2j+1) \frac{1}{\mu^{p(3 - 10 \frac{1}{q})}} \phi(j)^{-\frac{p}{q}} \, d\mu \right)^{\frac{1}{p}}.
\end{equation}

To compare the divergences on both sides of the inequality at the same leading order \( \sim \frac{1}{t} \) as \( t \to 0^- \), we choose \( q = 5 \), which satisfies the condition \( q > \frac{9}{2} \). This condition would not have been met without taking sufficiently many derivatives of \( M_2 \). Furthermore, the contribution from the \( c_2 \) term scales as \( c_2 (-t)^{-4/5} \) and is therefore subleading; it can be safely neglected in our analysis.

Regarding the parameter \( r \), which characterizes the size of the arc in the dispersion relation, the bound is optimized by choosing the maximal allowed value. Therefore, we set \( r = \Lambda^2 \), where \( \Lambda \) denotes the UV cutoff scale. In addition, we fix \( j_0 = 8 \), and choose the optimization parameter \( a = 2^{-\frac{1}{p}} \), which saturates the allowed upper bound discussed earlier. With these choices, the inequality becomes:

\begin{equation}
\frac{2^{- \frac{4}{5}}}{2 M_{\text{P}}^2} < 
\left( \frac{g_2^2}{240\pi^2} \right)^{\frac{1}{5}} 
\left( 
16 \sum_{j = 8}^{\infty} \int_{\Lambda^2}^\infty 
(2j + 1) \, \frac{ \phi(j)^{- \frac{1}{4}} }{ \mu^{5/4} } \, \mathrm{d}\mu 
\right)^{\frac{4}{5}}.
\end{equation}

Finally, we define the dimensionless coupling 
\( \tilde{g}_2 \equiv g_2 \Lambda^4 \). 
After performing the integral and summation explicitly, we arrive at the following bound:
\begin{equation}
   \frac{\Lambda}{M_{\text{P}}} < 2.38\, \tilde{g}_2^{1/5}.
\end{equation}

\section{Hölder's inequalities beyond the forward limit}
\label{sec:smeared}
The bound derived in the previous section is valid strictly when $t\rightarrow 0$ because the Legendre polynomials are oscillating functions in the limit of large $j\gg \sqrt{(-t)/\mu}$. For this reason, in the inequalities that we are using, this limit has to be taken more carefully. A possible way of a more rigorous justification of the bound can be done with the use of dispersion relations integrated over a finite range of $t$ with a smooth function. In addition, there is such a choice of this function that the Legendre polynomials would transform into positive-definite functions. This allows the use of Hölder's inequalities beyond the forward limit, and taking this limit at the end of the computation in a controllable way, keeping all functions in the inequality positive-definite. The latter is not possible without smearing because Legendre polynomials are oscillating at large $j$.

\subsection{Smeared dispersion relations with positive-definite functions}

Integrating the dispersion relation \eqref{B} one can obtain ($t=-q^2$),
\begin{equation}
    \label{Bs}
   \int^{q_0}_{0} f(q) M_2(q)dq= \sum_{J=0}^{\infty}\int_r^{\infty}16\,(2j+1)\,{\rm Im\,}f_j(\mu)d\mu\int_0^{q_0} f(q)d q \left( \frac{1}{\mu^{3}}+ \frac{1}{(\mu-q^2)^{3}}\right)P_j\left(1-\frac{2 q^2}{\mu}\right).
\end{equation}
Hereafter we assume $q_0^2\ll r$, so we can use an expansion
\begin{equation}
\begin{split}
  & \int^{q_0}_{0} f(q) M_2(q)dq= \\
   &=\sum_{J=0}^{\infty}\int_r^{\infty}16\,(2j+1)\,{\rm Im\,}f_j(\mu)d\mu\int_0^{q_0} f(q)d q \left( \frac{2}{\mu ^3}+\frac{3 q^2}{\mu ^4}+\frac{6 q^4}{\mu ^5}+\dots\right)P_j\left(1-\frac{2 q^2}{\mu}\right).
   \end{split}
\end{equation}
The choice of the smearing function in general can be arbitrary, given that it is smooth and the integral is convergent. For simplicity, we choose the smearing function as:
\begin{equation}
    f(q)=\frac{q^{\gamma}(q_0-q)^{\alpha}}{q_0^{\gamma+\alpha+1}}\frac{\Gamma(2+\alpha+\gamma)}{\Gamma(1+\alpha)\Gamma(1+\gamma)}.
\end{equation}
which ensures normalization over the interval. This choice of smearing functions allows for analytic integration of Legendre polynomials in terms of hypergeometric functions,
\begin{equation}
    \int_0^{q_0} d q f(q)\, P_J\left(1-\frac{2q^2}{\mu }\right)=  \, _4F_3\left(-J,J+1,\frac{\gamma }{2}+\frac{1}{2},\frac{\gamma }{2}+1;1,\frac{\alpha
   }{2}+\frac{\gamma }{2}+1,\frac{\alpha }{2}+\frac{\gamma
   }{2}+\frac{3}{2};\frac{q_0^2}{\mu }\right).
\end{equation}
\begin{figure}[h]
    \centering
    \includegraphics[width=0.5\linewidth]{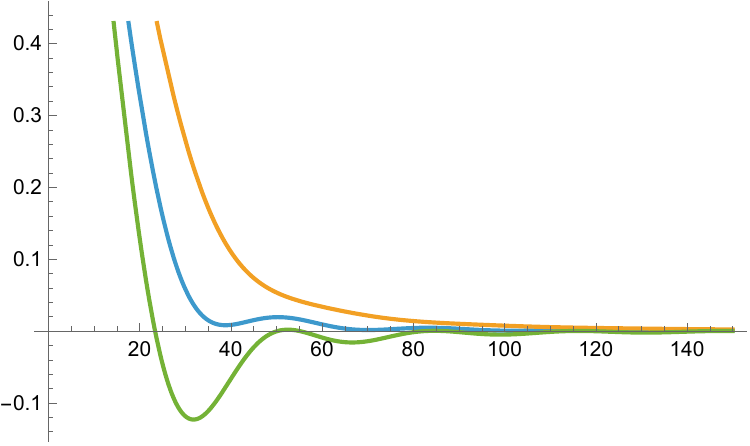}
    \caption{Hypergeometric functions for different smearing parameters.}
    \label{fig:refined}
\end{figure}
For the choice of $\gamma\leq 1$, $\alpha>3/2$, the corresponding function is always positive, thus making the sum and integral in the dispersion relation positive-definite. In this case, one can construct Hölder's inequalities, which are valid for finite $q_0$ if the right-hand side of them is properly chosen. Recall that on the right-hand side, one can use different smearing parameters, adjusted in such a way that the functions are also positive-definite for large $j$.

The method of the use of integral inequalities described in the previous sections can be straightforwardly generalized to arbitrary positive-definite functions. In order to make use of an inequality of this type,
\begin{equation}
    \int f g \,d\mu\leq\left( \int \left( f\right) ^{q}\,d\mu\right) ^{\frac{1}{q} }\left( \int \left( g\right) ^{p}\,d\mu\right) ^{\frac{1}{p}},\quad \frac{1}{p}+\frac{1}{q}=1,
\end{equation}
we can choose
\begin{equation}
\begin{split}
    &f g=({\rm Im\,}f_j)^{1/q}F(\mu,j),\\
    &f^q={\rm Im\,}f_j\, F_1(\mu,j),\\
    &g^p=F(\mu,j)^p(F_1(\mu,j))^{-p/q}.
\end{split}
\end{equation}
If we arrange the graviton pole part to be on the left side of an inequality, we take
\begin{equation}
\begin{split}
    &F(\mu,j)=\frac{\sqrt{\pi } (\gamma -1) \gamma  (2 j+1) 2^{-\alpha -\gamma +4} \Gamma (\alpha +\gamma )}{\mu ^3}\times\\
   \,&\times \,_4\tilde{F}_3\left(-j,j+1,\frac{\gamma +1}{2},\frac{\gamma +2}{2};1,\frac{1}{2} (\alpha
   +\gamma +2),\frac{1}{2} (\alpha +\gamma +3);\frac{q_0^2}{\mu }\right)+q_0^2(\dots)+\dots.
   \end{split}
\end{equation}
We found that the terms with higher powers of $q_0$ are not leading to large corrections to our results. This is expected, as we take the limit $q_0\rightarrow 0$ at the end. In this case, we have \footnote{The factor $16(2j+1)$ appears in the denominator because in \eqref{brackets} it is included in the bracket definition, while here we put it inside the definition of functions $F,~F_1$.}
\begin{equation}
    \frac{a}{2 M_{\text{P}}^2q_0^2}=\left[\frac{aF(\mu,j)}{16(2j+1)}\right]\leq\left[\frac{a|F(\mu,j)|}{16(2j+1)}\right]\leq \sum \int ({\rm Im\,}f_j)^{1/q}|F(\mu,j)|.
\end{equation}
In four dimensions, the function $F(\mu, j)$ is positive-definite everywhere only for $\gamma\leq 1$, which cannot be used for smearing of the graviton pole, as the integral would be divergent. Instead, we will be using $\gamma>1$ and an absolute value of $|F(\mu,j)|$ (obviously, $F(\mu,j)\leq|F(\mu,j)|$). As we will see later, this would make the bound a bit weaker, but would not invalidate it.

Since we now target getting the larger power of $1/q$, we take $q=2$. Thus, we need a combination that has the scaling of $1/q_0^4$ after smearing, or $1/t^2$. In this case, we can get the cancellation of the leading terms at $q_0\rightarrow 0$, and the limit of the inequality at $q_0\rightarrow 0$, which doesn't depend on the choice of $q_0$. The term behaving as $1/t^2$ can be obtained after taking five $t$-derivatives of $(-t)\,M_2$,
\begin{equation}
D_5 M_2=-\frac{\partial^5 (-t\,M_2)}{\partial t^5}=\frac{6 b_2}{t^2}+O(1/t).
\end{equation}
After smearing with the parameters $\gamma_1>3$ (required by the convergence) and $\alpha_1$ one can obtain
\begin{equation}
\overline{D_5 M_2}=    \frac{6 b_2}{q_0^4}\,\frac{ \Gamma \left(\gamma _1-3\right) \Gamma \left(\alpha _1+\gamma _1+2\right)}{
   \Gamma \left(\gamma _1+1\right) \Gamma \left(\alpha _1+\gamma _1-2\right)}.
\end{equation}
Thus, we can choose
\begin{equation}
    \frac{b_2}{q_0^4}=[F_1(\mu,j)],
\end{equation}
where the corresponding function is a combination of regularized hypergeometric functions, which is to be obtained as,
\begin{equation}
    F_1(\mu,j)=\frac{ 
   \Gamma \left(\gamma _1+1\right) \Gamma \left(\alpha _1+\gamma _1-2\right)}{\Gamma \left(\gamma _1-3\right) \Gamma \left(\alpha _1+\gamma _1+2\right)}\int_0^{q_0} f(q)d q \frac{\partial^5}{(\partial q^2)^5}\left(\frac{q^2}{6}\left( \frac{1}{\mu^{3}}+ \frac{1}{(\mu-q^2)^{3}}\right)P_j\left(1-\frac{2 q^2}{\mu}\right)\right).
\end{equation}
Here, the smearing function is parametrized by the powers $\gamma_1$ and $\alpha_1$. 

In the limit $q_0\rightarrow 0$, we have (this is valid only for small $j$),
\begin{equation}
\label{F1_series}
\begin{split}
  &F_1(\mu,j)=  \frac{10 \left(\gamma _1-3\right) \left(\gamma _1-2\right) \left(\gamma _1-1\right) \gamma _1
   (2 j+1) }{9 \mu ^7 \left(\alpha _1+\gamma _1-2\right)
   \left(\alpha _1+\gamma _1-1\right) \left(\alpha _1+\gamma _1\right) \left(\alpha _1+\gamma
   _1+1\right)}\times\\
   &\times\left((j-3) j (j+1) (j+4) \left(j (j+1)\left(j^2+j-32\right)+348\right)+4320\right)+q_0^2(\dots)+\dots.
   \end{split}
\end{equation}

Finally, dropping out the terms finite at $q_0\rightarrow 0$, such as $[F_1(\mu,2)]$ and $[F_1(\mu,0)]$ (in these points the function $F_1$ is negative), we have 
\begin{equation}
\label{bound}
    \frac{a}{2 M_{\text{P}}^2 q_0^2}<\left(\frac{b_2}{q_0^4}\right)^{1/2}\left(\sum_{j=4}^{\infty}\int_r^{\infty}d\mu F(\mu,j)^{2}(F_1(\mu,j))^{-1}\right)^{1/2}.
\end{equation}
Although the sum in $j$ is convergent, the integration with respect to $\mu$ would still lead to the divergence at infinity. Does it mean that the corresponding bound doesn't exist? 

In general, we could also have the parts of the $\mu$-integral which cannot lead to IR singularities in $M_2$ and $D_5 M_2$. If this is the case, we can use the inequality, taking the upper limit of the $\mu$-integral depending on $j$. In the next section, we explore such a possibility.

\subsection{The improved bound}

So far, we made an attempt to get a constraint comparing the sums and integrals where both upper limits are infinite,
\begin{equation}
    \frac{a}{2 M_{\text{P}}^2q_0^2}=\sum_{j=4}^{\infty}\int_r^{\infty}d \mu F(\mu,j)\leq\sum_{j=4}^{\infty}\int_r^{\infty}d \mu |F(\mu,j)|,\quad \frac{b_2}{q_0^4}=\sum_{j=4}^{\infty}\int_r^{\infty}d \mu F_1(\mu,j).
\end{equation}
However, one can use instead an integration with $j$-dependent upper limit,
\begin{equation}
    \frac{a}{2 M_{\text{P}}^2q_0^2}=\sum_{j=4}^{\infty}\int_r^{R(j)}d \mu F(\mu,j)+O(q_0^0),
\end{equation}
\begin{equation}
    \frac{b_2}{q_0^4}=\sum_{j=4}^{\infty}\int_r^{R(j)}d \mu F_1(\mu,j)+O(q_0^0).
\end{equation}
This representation works only if the remaining part of the integral between $R(j)$ and infinity provides only a finite contribution in the limit $q_0\rightarrow 0$. In other words, we can omit the subleading contributions in the limit $q_0\rightarrow 0$. We can estimate the tail of such integrals given the full unitarity condition ${\rm Im}\, f_j<2$. At large values of $\mu$ we can also use the low-spin expansion of $F,~F_1$, which is similar to \eqref{F1_series} and valid for $J^2 q_0/\mu\ll 1$.

Infrared $q_0\rightarrow 0$ singularities are reproduced from quite specific contributions corresponding to large $j$ and intermediate scales of $\mu$ which depend on $j$ (or impact parameter), see \cite{Haring:2024wyz} for the comprehensive study of the graviton pole case. In this example, an infinite part of the $\mu$ integral leads to a finite contribution. Indeed, let's assume that $R(j)=R_0 j^{\eta}$ and find the conditions on $\eta$ providing the convergent integrals $\int_{R(j)}^{\infty}$ of the functions $F$ and $F_1$. The estimates of the corresponding integrals for ${\rm Im\,}f_j=2$ are,
\begin{equation}
   F(\mu,j):~~ \sum_{j=4}^{\infty}\int_{R_0 j^{\eta}}^{\infty}d\mu\frac{j}{\mu^3}\sim\sum_{j=4}^{\infty}\frac{j^{1-2\eta}}{R_0^2},
\end{equation}
\begin{equation}
   F_1(\mu,j):~~ \sum_{j=4}^{\infty}\int_{R_0 j^{\eta}}^{\infty}d\mu\frac{j^9}{\mu^7}\sim\sum_{j=4}^{\infty}\frac{j^{9-6\eta}}{R_0^6}.
\end{equation}
The first sum converges for $\eta>1$, while the second is finite for $\eta>5/3$\footnote{Recall that if both sum and integral do converge, the order of these two operations (summation and integration) doesn't matter.}. It is also worth mentioning that the condition $j^2 q_0^2/\mu\ll 1$ can be used at the lower limit of these integrals $\mu=R_0\j^{\eta}$ for arbitrarily large $j$ if $\eta\geq 2$. 

The requirement of having a finite sum on the right-hand side of \eqref{bound} would provide an upper bound on $\eta$. Indeed,
\begin{equation}
    \sum_{j=4}^{\infty}\int_{r}^{R_0 j^{\eta}}d \mu F(\mu,j)^{2}(F_1(\mu,j))^{-1}\sim  \sum_{j=4}^{\infty}\int_{r}^{R_0 j^{\eta}}\frac{j^2}{\mu^6}\frac{\mu^7}{j^9}d \mu\sim \sum_{j=4}^{\infty}\left.\frac{\mu^2}{j^7}\right|_{\mu=r}^{\mu=R_0 j^{\eta}}=\sum_{j=4}^{\infty}R_0^2 j^{2\eta-7}.
\end{equation}
This series is convergent for $\eta<3$. Thus, taking $2\leq\eta<3$ would make the bound \eqref{bound} convergent while the corrections to the IR quantities would still be finite in the limit $q_0\rightarrow 0$, so they can be safely neglected. 

Finally, we can get a bound in the following form,
\begin{equation}
\label{bound-finite}
    \frac{1}{2\sqrt{2} M_{\text{P}}^2 q_0^2}+O(q_0^0)<\left(\frac{b_2}{q_0^4}+O(q_0^0)\right)^{1/2}\left(\sum_{j=4}^{\infty}\int_r^{R_0 j^{\eta}}d\mu F(\mu,j)^{2}(F_1(\mu,j))^{-1}\right)^{1/2},\quad q_0\rightarrow 0.
\end{equation}
Here we choose the optimal value of $a=2^{-1/2}$, and take $R_0=r$, $\eta=2$, $q_0=0.02\sqrt{r}$. We perform a numerical evaluation of the sum and integral,
\begin{equation}
  S= \left(\sum_{j=4}^{\infty}S_j \right)^{1/2}= \left(\sum_{j=4}^{\infty}\int_r^{R_0 j^{\eta}}d\mu F(\mu,j)^{2}(F_1(\mu,j))^{-1}\right)^{1/2}
\end{equation}

We checked the following values of the smearing parameters,
\begin{equation}
    \gamma=1.1,~\alpha=2,~\gamma_1=3.1, ~\alpha_1=1, ~~S=0.939,
\end{equation}
\begin{equation}
    \gamma=1.05,~\alpha=3/2,~\gamma_1=8.1, ~\alpha_1=6, ~~S=0.198.
\end{equation}
We found that the bound is slightly improved for larger values of $\gamma_1$, but this parameter cannot be increased to large values because it would make the function $F_1$ non-positive for large $j$ ( we found that this is the case for $\gamma_1\gtrsim 9$). Recall that we can still allow for having negative values $F$ because we can take $|F|$ in the left-hand side of the inequality, while it is not possible for the right-hand side. For this reason, the value of $S$ cannot be made significantly smaller than $0.1$, although we leave the accurate optimization of this bound for future work. 

\begin{figure}[h]
    \centering
    \includegraphics[width=0.4\linewidth]{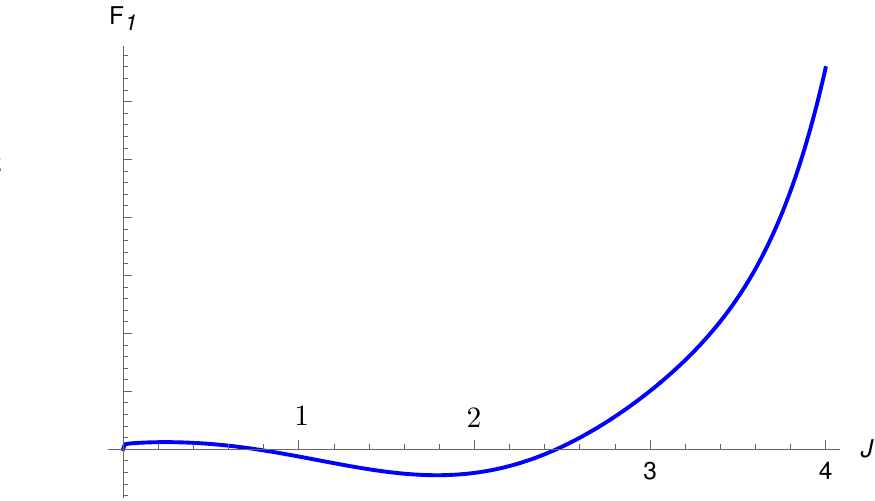}\qquad\qquad\qquad
    \includegraphics[width=0.4\linewidth]{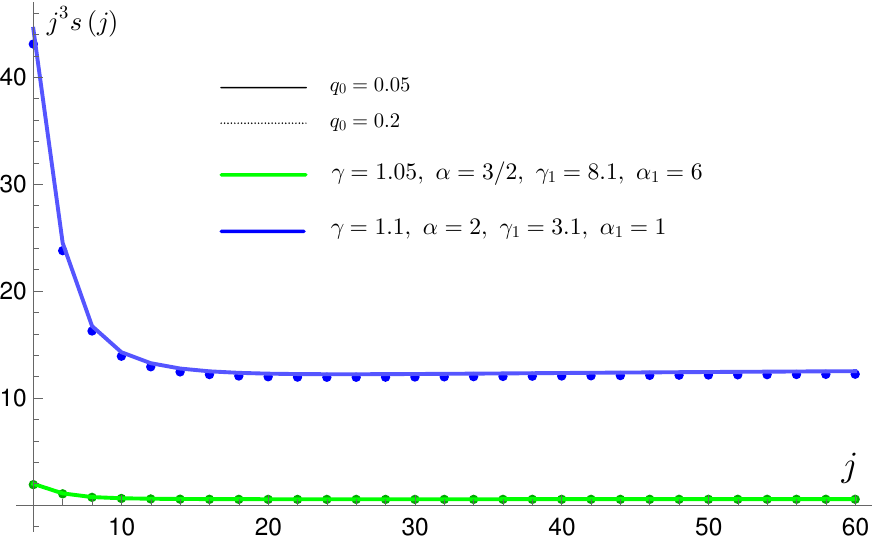}
    \caption{A sketch of the hypergeometric function $ F_1(\mu,j)$ for small $j$ (left plot) and  values of $j^3 S_j$ for different smearing parameters and $\eta=2$ (right plot).}
    \label{fig:sums}
\end{figure}

The terms in the sum $S_j$ behave as $1/j^3$ for $\eta=1$, as it is shown in \ref{fig:sums}, so the corresponding series shows quick convergence for large $j$. The result also does not visibly change when $q_0\rightarrow 0$, such that $q_0=0.05$ is enough to reproduce the limit to a good precision.

\begin{figure}[h]
    \centering
    \includegraphics[width=0.4\linewidth]{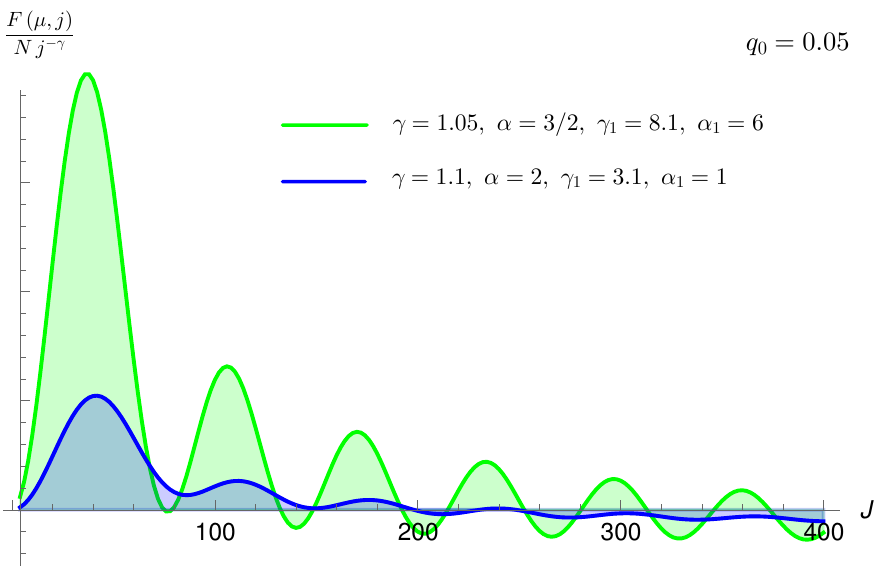}\qquad
    \includegraphics[width=0.4\linewidth]{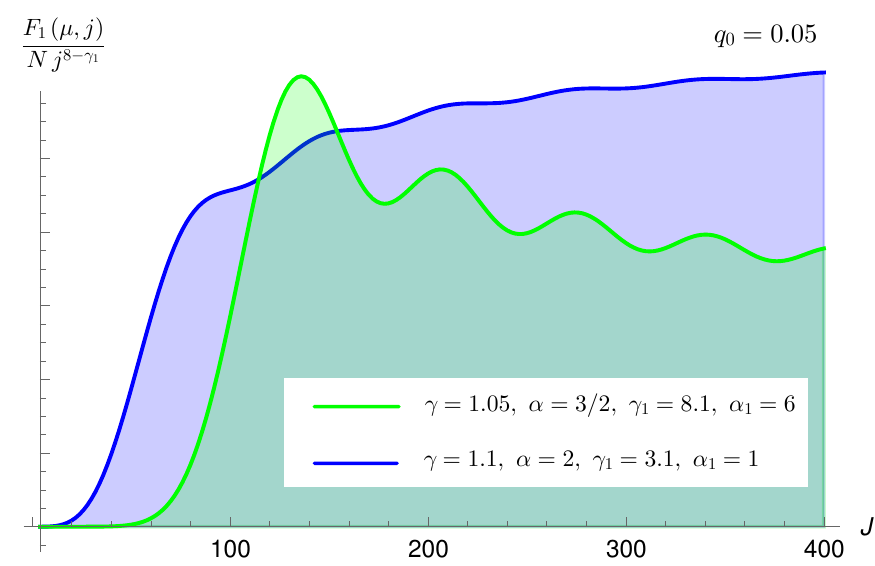}
    \caption{Hypergeometric functions $ F_1(\mu,j)$ (left plot) and $F(\mu,j)$ (right plot) for different smearing parameters. The normalization factor $N$ is adjusted such that the values of the plotted functions are of order 1.}
    \label{fig:F1}
\end{figure}

Summarizing, we obtained
\begin{equation}
    \frac{\Lambda^2}{M_P^2}<0.0115 |\tilde{g}_2|.
\end{equation}

In this work, we aim to show the existence of a bound on gravitational coupling with $S=O(1)$, leaving the question of how it can be further optimized for future study. In principle, a lot of different inequalities of the discussed type can be written, which leads to various powers $b_2$ in the right-hand side. However, allowing for more $t$-derivatives on both sides has to be treated more carefully because of the presence of more non-analytic terms related to the loops of gravitons. One has to be sure that the right-hand side is the dominant contribution in the forward limit, even if loops are taken into account, which is very hard to estimate for higher-order terms. In the next section, we provide arguments showing that the current bound is stable against the inclusion of graviton loops.

\subsection{A comment on corrections from graviton loops}

Graviton loops in four dimensions would affect the structure of the arc integral as\footnote{Here we included only one-loop non-analytic terms, see the results presented in \cite{Beadle:2025cdx,Chang:2025cxc}.},
\begin{equation}    
    M_2= \frac{A}{t}+B\, \log{(-t)} + C\, t \log{(-t)}+  D\, t^2 \log{(-t)}+\dots .
\end{equation}
The combination we used throughout this paper,
\begin{equation}
     \frac{\partial}{\partial t}(t M_2)= B \log (-t)+B+C t+2 C t \log (-t)+D t^2+3 D t^2 \log (-t),
\end{equation}
eliminates the graviton pole; however, there are more singular contributions compared to $t^2 \log{(-t)}$. They can also be eliminated similarly,
\begin{equation}
\label{no logs}
  \frac{\partial^3}{\partial t^3}\left(t\frac{\partial^2}{\partial t^2} \left(t \frac{\partial^2}{\partial t^2}(t M_2)\right)\right)=-12\frac{D}{t^2}.
\end{equation}
This combination picks the $t^2 \log{(-t)}$ term from the expansion of $M_2$, while for the polynomial it $t$ it has the same form as the expansion used in the previous section. Thus, if we use the expression \eqref{no logs} instead of $\overline{D_5 M_2}$, we get the same bound. The only difference caused by loops would be related to the term of the form $t^2 \log{(-t)}$ coming from graviton loops, which mixes with the same non-analytic structure from the scalar loop. In this case, we have
\begin{equation}
    \frac{\Lambda^2}{M_P^2}<0.0115 \sqrt{\tilde{g}_2^2 +O(1/M_P^6)}.
\end{equation}
By power-counting in four dimensions, we can find that the term of the form $s^2 t^2 \log {(-t)}$ in the amplitude can be generated from GR action only at two loops, hence it should be suppressed at least by an extra $M_P^2$. This makes our bound non-trivial and stable with respect to the presence of graviton loops.

We expect that non-analytic terms from higher loops (being always power series of logarithms) can be eliminated in a similar way as logs in \eqref{no logs}. The structure $s^2 t^2 \log {(-t)}$ can also emerge at one-loop from mixed couplings between EFT of scalar and EFT of gravity; however, we still expect that it would lead to a non-trivial bound stating that $g_2$ coupling is bounded from below in the presence of gravity. It is easy to understand from the point of view that a scalar field coupled to gravity cannot remain a free theory. Graviton loops would generate $(\partial \phi)^4$ counterterm even if one started from the scalar theory minimally coupled to gravity. Our bound states that this counterterm cannot be renormalized to zero value. Instead, it should be of order $1/M_P^4$, which is naturally expected from power-counting arguments.

\section{Conclusions}
\label{sec:conclusions}
In this work, we derived a weak gravity bound by applying the Hölder inequality to the integral expressions arising from the positivity and full unitarity constraints in four dimensions. We derived a series of inequalities that provide upper bounds on the Planck mass in terms of physical quantities such as the dimensionless coupling \( \tilde{g}_2 \) and the cutoff scale $\Lambda$ of the scalar EFT.

As an additional assumption, compared to several recent papers on finite$-t$ gravitational positivity bounds, we required full unitarity of partial waves in four dimensions, which means that we bridge the EFT to the UV completion, which is formulated as a unitary theory in four dimensions. In addition, the technique we used is based on taking the forward limit, comparing the IR singularities of different nature by the use of Hölder's inequalities. In fact, we are constraining $\beta$-functions of EFT couplings, which is different from the previous efforts focused on constraining the couplings. For these reasons, our result could not be reproduced from the techniques based on finite $t$ dispersion relations used so far. In this work, we mainly focus on proving the existence of the weak gravity bound, leaving the proper optimization of it, together with the other constraints, for future study.

The weak gravity bound provides valuable insights into the structure of quantum gravity theories and offers a consistent way for coupling them to the scalar field. The bound we obtained tells us that if a free scalar theory is coupled to gravity, it requires the scalar self-coupling to be introduced for consistency. Intuitively, one can expect that graviton loops would generate such a coupling suppressed at least by the Planck scale. Our result confirms this intuition and forbids the possibility of keeping the scalar field without a four-derivative self-interaction term. This conclusion can have far-reaching implications for the early Universe inflation models because the derivative coupling could lead to a specific shape of non-gaussianities in the power spectrum. In addition, this coupling could play an important role after inflation during the preheating stage. Another implication of our bound is yet another way to state that the Galileon symmetry of the scalar field (corresponding to $g_2=0$) is incompatible with unitary UV completion even in the presence of gravity. 

The techniques developed here are applicable to a broader class of models, including particles with spin, where positivity constraints play a central role. Given that there are similar recurrence relations for $D$-dimensional Legendre polynomials, it is also straightforward to extend our methods to the theories in any number of dimensions, including the case when four-dimensional EFT is linked to $D$-dimensional UV completion. Future work could involve further refinement of the bound by considering additional physical constraints or exploring alternative methods for handling and regularizing the infinite sums and integrals in the calculations.


\subsection*{Acknowledgements}

A.~T. thanks Ivano Basile, Francesco Riva, Laurentiu Rodina, Shilin Wan, Alexander Zhiboedov, Shuang-Yong Zhou for several illuminating discussions, comments, and reasonable criticism. The work of A.~T. was supported by the National Natural Science Foundation of China (NSFC) under Grant No. 1234710.

\bibliographystyle{utphys}
\bibliography{References}

\providecommand{\href}[2]{#2}\begingroup\raggedright\begin{thebibliography}{100}

\bibitem{Arkani-Hamed:2006emk}
N.~Arkani-Hamed, L.~Motl, A.~Nicolis, and C.~Vafa, ``{The String landscape, black holes and gravity as the weakest force},'' \href{http://dx.doi.org/10.1088/1126-6708/2007/06/060}{{\em JHEP} {\bfseries 06} (2007) 060}, \href{http://arxiv.org/abs/hep-th/0601001}{{\ttfamily arXiv:hep-th/0601001}}.

\bibitem{2401.14449}
B.~Heidenreich and M.~Lotito, ``{Proving the Weak Gravity Conjecture in Perturbative String Theory, Part I: The Bosonic String},'' \href{http://arxiv.org/abs/2401.14449}{{\ttfamily arXiv:2401.14449 [hep-th]}}.

\bibitem{1903.06239}
E.~Palti, ``{The Swampland: Introduction and Review},'' \href{http://dx.doi.org/10.1002/prop.201900037}{{\em Fortsch. Phys.} {\bfseries 67} no.~6, (2019) 1900037}, \href{http://arxiv.org/abs/1903.06239}{{\ttfamily arXiv:1903.06239 [hep-th]}}.

\bibitem{2201.08380}
D.~Harlow, B.~Heidenreich, M.~Reece, and T.~Rudelius, ``{Weak gravity conjecture},'' \href{http://dx.doi.org/10.1103/RevModPhys.95.035003}{{\em Rev. Mod. Phys.} {\bfseries 95} no.~3, (2023) 035003}, \href{http://arxiv.org/abs/2201.08380}{{\ttfamily arXiv:2201.08380 [hep-th]}}.

\bibitem{2409.10003}
A.~Castellano, {\em {The Quantum Gravity Scale and the Swampland}}.
\newblock PhD thesis, U. Autonoma, Madrid (main), 2024.
\newblock \href{http://arxiv.org/abs/2409.10003}{{\ttfamily arXiv:2409.10003 [hep-th]}}.

\bibitem{Bastian:2020egp}
B.~Bastian, T.~W. Grimm, and D.~van~de Heisteeg, ``{Weak gravity bounds in asymptotic string compactifications},'' \href{http://dx.doi.org/10.1007/JHEP06(2021)162}{{\em JHEP} {\bfseries 06} (2021) 162}, \href{http://arxiv.org/abs/2011.08854}{{\ttfamily arXiv:2011.08854 [hep-th]}}.

\bibitem{Cheung:2018cwt}
C.~Cheung, J.~Liu, and G.~N. Remmen, ``{Proof of the Weak Gravity Conjecture from Black Hole Entropy},'' \href{http://dx.doi.org/10.1007/JHEP10(2018)004}{{\em JHEP} {\bfseries 10} (2018) 004}, \href{http://arxiv.org/abs/1801.08546}{{\ttfamily arXiv:1801.08546 [hep-th]}}.

\bibitem{Cottrell:2016bty}
G.~Shiu, P.~Soler, and W.~Cottrell, ``{Weak Gravity Conjecture and extremal black holes},'' \href{http://dx.doi.org/10.1007/s11433-019-9406-2}{{\em Sci. China Phys. Mech. Astron.} {\bfseries 62} no.~11, (2019) 110412}, \href{http://arxiv.org/abs/1611.06270}{{\ttfamily arXiv:1611.06270 [hep-th]}}.

\bibitem{Hebecker:2017uix}
A.~Hebecker and P.~Soler, ``{The Weak Gravity Conjecture and the Axionic Black Hole Paradox},'' \href{http://dx.doi.org/10.1007/JHEP09(2017)036}{{\em JHEP} {\bfseries 09} (2017) 036}, \href{http://arxiv.org/abs/1702.06130}{{\ttfamily arXiv:1702.06130 [hep-th]}}.

\bibitem{Abe:2023anf}
Y.~Abe, T.~Noumi, and K.~Yoshimura, ``{Black hole extremality in nonlinear electrodynamics: a lesson for weak gravity and Festina Lente bounds},'' \href{http://dx.doi.org/10.1007/JHEP09(2023)024}{{\em JHEP} {\bfseries 09} (2023) 024}, \href{http://arxiv.org/abs/2305.17062}{{\ttfamily arXiv:2305.17062 [hep-th]}}.

\bibitem{DeLuca:2022tkm}
V.~De~Luca, J.~Khoury, and S.~S.~C. Wong, ``{Implications of the weak gravity conjecture for tidal Love numbers of black holes},'' \href{http://dx.doi.org/10.1103/PhysRevD.108.044066}{{\em Phys. Rev. D} {\bfseries 108} no.~4, (2023) 044066}, \href{http://arxiv.org/abs/2211.14325}{{\ttfamily arXiv:2211.14325 [hep-th]}}.

\bibitem{Barbosa:2025uau}
S.~Barbosa, P.~Brax, S.~Fichet, and L.~de~Souza, ``{Running Love Numbers and the Effective Field Theory of Gravity},'' \href{http://arxiv.org/abs/2501.18684}{{\ttfamily arXiv:2501.18684 [hep-th]}}.

\bibitem{Cao:2022ajt}
Q.-H. Cao, N.~Kan, and D.~Ueda, ``{Effective field theory in light of relative entropy},'' \href{http://dx.doi.org/10.1007/JHEP07(2023)111}{{\em JHEP} {\bfseries 07} (2023) 111}, \href{http://arxiv.org/abs/2211.08065}{{\ttfamily arXiv:2211.08065 [hep-th]}}.

\bibitem{Cao:2022iqh}
Q.-H. Cao and D.~Ueda, ``{Entropy constraints on effective field theory},'' \href{http://dx.doi.org/10.1103/PhysRevD.108.025011}{{\em Phys. Rev. D} {\bfseries 108} no.~2, (2023) 025011}, \href{http://arxiv.org/abs/2201.00931}{{\ttfamily arXiv:2201.00931 [hep-th]}}.

\bibitem{Aalsma:2019ryi}
L.~Aalsma, A.~Cole, and G.~Shiu, ``{Weak Gravity Conjecture, Black Hole Entropy, and Modular Invariance},'' \href{http://dx.doi.org/10.1007/JHEP08(2019)022}{{\em JHEP} {\bfseries 08} (2019) 022}, \href{http://arxiv.org/abs/1905.06956}{{\ttfamily arXiv:1905.06956 [hep-th]}}.

\bibitem{Bellazzini:2019xts}
B.~Bellazzini, M.~Lewandowski, and J.~Serra, ``{Positivity of Amplitudes, Weak Gravity Conjecture, and Modified Gravity},'' \href{http://dx.doi.org/10.1103/PhysRevLett.123.251103}{{\em Phys. Rev. Lett.} {\bfseries 123} no.~25, (2019) 251103}, \href{http://arxiv.org/abs/1902.03250}{{\ttfamily arXiv:1902.03250 [hep-th]}}.

\bibitem{Hamada:2018dde}
Y.~Hamada, T.~Noumi, and G.~Shiu, ``{Weak Gravity Conjecture from Unitarity and Causality},'' \href{http://dx.doi.org/10.1103/PhysRevLett.123.051601}{{\em Phys. Rev. Lett.} {\bfseries 123} no.~5, (2019) 051601}, \href{http://arxiv.org/abs/1810.03637}{{\ttfamily arXiv:1810.03637 [hep-th]}}.

\bibitem{Arkani-Hamed:2021ajd}
N.~Arkani-Hamed, Y.-t. Huang, J.-Y. Liu, and G.~N. Remmen, ``{Causality, unitarity, and the weak gravity conjecture},'' \href{http://dx.doi.org/10.1007/JHEP03(2022)083}{{\em JHEP} {\bfseries 03} (2022) 083}, \href{http://arxiv.org/abs/2109.13937}{{\ttfamily arXiv:2109.13937 [hep-th]}}.

\bibitem{Henriksson:2021ymi}
J.~Henriksson, B.~McPeak, F.~Russo, and A.~Vichi, ``{Rigorous bounds on light-by-light scattering},'' \href{http://dx.doi.org/10.1007/JHEP06(2022)158}{{\em JHEP} {\bfseries 06} (2022) 158}, \href{http://arxiv.org/abs/2107.13009}{{\ttfamily arXiv:2107.13009 [hep-th]}}.

\bibitem{Alberte:2020bdz}
L.~Alberte, C.~de~Rham, S.~Jaitly, and A.~J. Tolley, ``{QED positivity bounds},'' \href{http://dx.doi.org/10.1103/PhysRevD.103.125020}{{\em Phys. Rev. D} {\bfseries 103} no.~12, (2021) 125020}, \href{http://arxiv.org/abs/2012.05798}{{\ttfamily arXiv:2012.05798 [hep-th]}}.

\bibitem{Nakayama:2015hga}
Y.~Nakayama and Y.~Nomura, ``{Weak gravity conjecture in the AdS/CFT correspondence},'' \href{http://dx.doi.org/10.1103/PhysRevD.92.126006}{{\em Phys. Rev. D} {\bfseries 92} no.~12, (2015) 126006}, \href{http://arxiv.org/abs/1509.01647}{{\ttfamily arXiv:1509.01647 [hep-th]}}.

\bibitem{Harlow:2015lma}
D.~Harlow, ``{Wormholes, Emergent Gauge Fields, and the Weak Gravity Conjecture},'' \href{http://dx.doi.org/10.1007/JHEP01(2016)122}{{\em JHEP} {\bfseries 01} (2016) 122}, \href{http://arxiv.org/abs/1510.07911}{{\ttfamily arXiv:1510.07911 [hep-th]}}.

\bibitem{Benjamin:2016fhe}
N.~Benjamin, E.~Dyer, A.~L. Fitzpatrick, and S.~Kachru, ``{Universal Bounds on Charged States in 2d CFT and 3d Gravity},'' \href{http://dx.doi.org/10.1007/JHEP08(2016)041}{{\em JHEP} {\bfseries 08} (2016) 041}, \href{http://arxiv.org/abs/1603.09745}{{\ttfamily arXiv:1603.09745 [hep-th]}}.

\bibitem{Montero:2016tif}
M.~Montero, G.~Shiu, and P.~Soler, ``{The Weak Gravity Conjecture in three dimensions},'' \href{http://dx.doi.org/10.1007/JHEP10(2016)159}{{\em JHEP} {\bfseries 10} (2016) 159}, \href{http://arxiv.org/abs/1606.08438}{{\ttfamily arXiv:1606.08438 [hep-th]}}.

\bibitem{Horowitz:2016ezu}
G.~T. Horowitz, J.~E. Santos, and B.~Way, ``{Evidence for an Electrifying Violation of Cosmic Censorship},'' \href{http://dx.doi.org/10.1088/0264-9381/33/19/195007}{{\em Class. Quant. Grav.} {\bfseries 33} no.~19, (2016) 195007}, \href{http://arxiv.org/abs/1604.06465}{{\ttfamily arXiv:1604.06465 [hep-th]}}.

\bibitem{Crisford:2017gsb}
T.~Crisford, G.~T. Horowitz, and J.~E. Santos, ``{Testing the Weak Gravity - Cosmic Censorship Connection},'' \href{http://dx.doi.org/10.1103/PhysRevD.97.066005}{{\em Phys. Rev. D} {\bfseries 97} no.~6, (2018) 066005}, \href{http://arxiv.org/abs/1709.07880}{{\ttfamily arXiv:1709.07880 [hep-th]}}.

\bibitem{Yu:2018eqq}
T.-Y. Yu and W.-Y. Wen, ``{Cosmic censorship and Weak Gravity Conjecture in the Einstein\textendash{}Maxwell-dilaton theory},'' \href{http://dx.doi.org/10.1016/j.physletb.2018.04.060}{{\em Phys. Lett. B} {\bfseries 781} (2018) 713--718}, \href{http://arxiv.org/abs/1803.07916}{{\ttfamily arXiv:1803.07916 [gr-qc]}}.

\bibitem{Brown:2015iha}
J.~Brown, W.~Cottrell, G.~Shiu, and P.~Soler, ``{Fencing in the Swampland: Quantum Gravity Constraints on Large Field Inflation},'' \href{http://dx.doi.org/10.1007/JHEP10(2015)023}{{\em JHEP} {\bfseries 10} (2015) 023}, \href{http://arxiv.org/abs/1503.04783}{{\ttfamily arXiv:1503.04783 [hep-th]}}.

\bibitem{Brown:2015lia}
J.~Brown, W.~Cottrell, G.~Shiu, and P.~Soler, ``{On Axionic Field Ranges, Loopholes and the Weak Gravity Conjecture},'' \href{http://dx.doi.org/10.1007/JHEP04(2016)017}{{\em JHEP} {\bfseries 04} (2016) 017}, \href{http://arxiv.org/abs/1504.00659}{{\ttfamily arXiv:1504.00659 [hep-th]}}.

\bibitem{Heidenreich:2015nta}
B.~Heidenreich, M.~Reece, and T.~Rudelius, ``{Sharpening the Weak Gravity Conjecture with Dimensional Reduction},'' \href{http://dx.doi.org/10.1007/JHEP02(2016)140}{{\em JHEP} {\bfseries 02} (2016) 140}, \href{http://arxiv.org/abs/1509.06374}{{\ttfamily arXiv:1509.06374 [hep-th]}}.

\bibitem{Heidenreich:2016aqi}
B.~Heidenreich, M.~Reece, and T.~Rudelius, ``{Evidence for a sublattice weak gravity conjecture},'' \href{http://dx.doi.org/10.1007/JHEP08(2017)025}{{\em JHEP} {\bfseries 08} (2017) 025}, \href{http://arxiv.org/abs/1606.08437}{{\ttfamily arXiv:1606.08437 [hep-th]}}.

\bibitem{Lee:2018urn}
S.-J. Lee, W.~Lerche, and T.~Weigand, ``{Tensionless Strings and the Weak Gravity Conjecture},'' \href{http://dx.doi.org/10.1007/JHEP10(2018)164}{{\em JHEP} {\bfseries 10} (2018) 164}, \href{http://arxiv.org/abs/1808.05958}{{\ttfamily arXiv:1808.05958 [hep-th]}}.

\bibitem{Cheung:2014ega}
C.~Cheung and G.~N. Remmen, ``{Infrared Consistency and the Weak Gravity Conjecture},'' \href{http://dx.doi.org/10.1007/JHEP12(2014)087}{{\em JHEP} {\bfseries 12} (2014) 087}, \href{http://arxiv.org/abs/1407.7865}{{\ttfamily arXiv:1407.7865 [hep-th]}}.

\bibitem{Andriolo:2018lvp}
S.~Andriolo, D.~Junghans, T.~Noumi, and G.~Shiu, ``{A Tower Weak Gravity Conjecture from Infrared Consistency},'' \href{http://dx.doi.org/10.1002/prop.201800020}{{\em Fortsch. Phys.} {\bfseries 66} no.~5, (2018) 1800020}, \href{http://arxiv.org/abs/1802.04287}{{\ttfamily arXiv:1802.04287 [hep-th]}}.

\bibitem{Bittar:2024xuc}
P.~Bittar, S.~Fichet, and L.~de~Souza, ``{Gravity-Induced Photon Interactions and Infrared Consistency in any Dimensions},'' \href{http://arxiv.org/abs/2404.07254}{{\ttfamily arXiv:2404.07254 [hep-th]}}.

\bibitem{Saraswat:2016eaz}
P.~Saraswat, ``{Weak gravity conjecture and effective field theory},'' \href{http://dx.doi.org/10.1103/PhysRevD.95.025013}{{\em Phys. Rev. D} {\bfseries 95} no.~2, (2017) 025013}, \href{http://arxiv.org/abs/1608.06951}{{\ttfamily arXiv:1608.06951 [hep-th]}}.

\bibitem{Guerrieri:2021ivu}
A.~Guerrieri, J.~Penedones, and P.~Vieira, ``{Where Is String Theory in the Space of Scattering Amplitudes?},'' \href{http://dx.doi.org/10.1103/PhysRevLett.127.081601}{{\em Phys. Rev. Lett.} {\bfseries 127} no.~8, (2021) 081601}, \href{http://arxiv.org/abs/2102.02847}{{\ttfamily arXiv:2102.02847 [hep-th]}}.

\bibitem{Pham:1985cr}
T.~N. Pham and T.~N. Truong, ``{Evaluation of the Derivative Quartic Terms of the Meson Chiral Lagrangian From Forward Dispersion Relation},'' \href{http://dx.doi.org/10.1103/PhysRevD.31.3027}{{\em Phys. Rev. D} {\bfseries 31} (1985) 3027}.

\bibitem{Pennington:1994kc}
M.~R. Pennington and J.~Portoles, ``{The Chiral Lagrangian parameters, l1, l2, are determined by the rho resonance},'' \href{http://dx.doi.org/10.1016/0370-2693(94)01551-M}{{\em Phys. Lett. B} {\bfseries 344} (1995) 399--406}, \href{http://arxiv.org/abs/hep-ph/9409426}{{\ttfamily arXiv:hep-ph/9409426}}.

\bibitem{Nicolis:2009qm}
A.~Nicolis, R.~Rattazzi, and E.~Trincherini, ``{Energy's and amplitudes' positivity},'' \href{http://dx.doi.org/10.1007/JHEP05(2010)095}{{\em JHEP} {\bfseries 05} (2010) 095}, \href{http://arxiv.org/abs/0912.4258}{{\ttfamily arXiv:0912.4258 [hep-th]}}. [Erratum: JHEP 11, 128 (2011)].

\bibitem{Komargodski:2011vj}
Z.~Komargodski and A.~Schwimmer, ``{On Renormalization Group Flows in Four Dimensions},'' \href{http://dx.doi.org/10.1007/JHEP12(2011)099}{{\em JHEP} {\bfseries 12} (2011) 099}, \href{http://arxiv.org/abs/1107.3987}{{\ttfamily arXiv:1107.3987 [hep-th]}}.

\bibitem{Remmen:2019cyz}
G.~N. Remmen and N.~L. Rodd, ``{Consistency of the Standard Model Effective Field Theory},'' \href{http://dx.doi.org/10.1007/JHEP12(2019)032}{{\em JHEP} {\bfseries 12} (2019) 032}, \href{http://arxiv.org/abs/1908.09845}{{\ttfamily arXiv:1908.09845 [hep-ph]}}.

\bibitem{Herrero-Valea:2019hde}
M.~Herrero-Valea, I.~Timiryasov, and A.~Tokareva, ``{To Positivity and Beyond, where Higgs-Dilaton Inflation has never gone before},'' \href{http://dx.doi.org/10.1088/1475-7516/2019/11/042}{{\em JCAP} {\bfseries 11} (2019) 042}, \href{http://arxiv.org/abs/1905.08816}{{\ttfamily arXiv:1905.08816 [hep-ph]}}.

\bibitem{Bellazzini:2017fep}
B.~Bellazzini, F.~Riva, J.~Serra, and F.~Sgarlata, ``{Beyond Positivity Bounds and the Fate of Massive Gravity},'' \href{http://dx.doi.org/10.1103/PhysRevLett.120.161101}{{\em Phys. Rev. Lett.} {\bfseries 120} no.~16, (2018) 161101}, \href{http://arxiv.org/abs/1710.02539}{{\ttfamily arXiv:1710.02539 [hep-th]}}.

\bibitem{deRham:2017avq}
C.~de~Rham, S.~Melville, A.~J. Tolley, and S.-Y. Zhou, ``{Positivity bounds for scalar field theories},'' \href{http://dx.doi.org/10.1103/PhysRevD.96.081702}{{\em Phys. Rev. D} {\bfseries 96} no.~8, (2017) 081702}, \href{http://arxiv.org/abs/1702.06134}{{\ttfamily arXiv:1702.06134 [hep-th]}}.

\bibitem{deRham:2017zjm}
C.~de~Rham, S.~Melville, A.~J. Tolley, and S.-Y. Zhou, ``{UV complete me: Positivity Bounds for Particles with Spin},'' \href{http://dx.doi.org/10.1007/JHEP03(2018)011}{{\em JHEP} {\bfseries 03} (2018) 011}, \href{http://arxiv.org/abs/1706.02712}{{\ttfamily arXiv:1706.02712 [hep-th]}}.

\bibitem{deRham:2017imi}
C.~de~Rham, S.~Melville, A.~J. Tolley, and S.-Y. Zhou, ``{Massive Galileon Positivity Bounds},'' \href{http://dx.doi.org/10.1007/JHEP09(2017)072}{{\em JHEP} {\bfseries 09} (2017) 072}, \href{http://arxiv.org/abs/1702.08577}{{\ttfamily arXiv:1702.08577 [hep-th]}}.

\bibitem{Wang:2020jxr}
Y.-J. Wang, F.-K. Guo, C.~Zhang, and S.-Y. Zhou, ``{Generalized positivity bounds on chiral perturbation theory},'' \href{http://dx.doi.org/10.1007/JHEP07(2020)214}{{\em JHEP} {\bfseries 07} (2020) 214}, \href{http://arxiv.org/abs/2004.03992}{{\ttfamily arXiv:2004.03992 [hep-ph]}}.

\bibitem{Tokuda:2020mlf}
J.~Tokuda, K.~Aoki, and S.~Hirano, ``{Gravitational positivity bounds},'' \href{http://dx.doi.org/10.1007/JHEP11(2020)054}{{\em JHEP} {\bfseries 11} (2020) 054}, \href{http://arxiv.org/abs/2007.15009}{{\ttfamily arXiv:2007.15009 [hep-th]}}.

\bibitem{Li:2021lpe}
X.~Li, H.~Xu, C.~Yang, C.~Zhang, and S.-Y. Zhou, ``{Positivity in Multifield Effective Field Theories},'' \href{http://dx.doi.org/10.1103/PhysRevLett.127.121601}{{\em Phys. Rev. Lett.} {\bfseries 127} no.~12, (2021) 121601}, \href{http://arxiv.org/abs/2101.01191}{{\ttfamily arXiv:2101.01191 [hep-ph]}}.

\bibitem{Caron-Huot:2021rmr}
S.~Caron-Huot, D.~Mazac, L.~Rastelli, and D.~Simmons-Duffin, ``{Sharp boundaries for the swampland},'' \href{http://dx.doi.org/10.1007/JHEP07(2021)110}{{\em JHEP} {\bfseries 07} (2021) 110}, \href{http://arxiv.org/abs/2102.08951}{{\ttfamily arXiv:2102.08951 [hep-th]}}.

\bibitem{Du:2021byy}
Z.-Z. Du, C.~Zhang, and S.-Y. Zhou, ``{Triple crossing positivity bounds for multi-field theories},'' \href{http://dx.doi.org/10.1007/JHEP12(2021)115}{{\em JHEP} {\bfseries 12} (2021) 115}, \href{http://arxiv.org/abs/2111.01169}{{\ttfamily arXiv:2111.01169 [hep-th]}}.

\bibitem{Bern:2021ppb}
Z.~Bern, D.~Kosmopoulos, and A.~Zhiboedov, ``{Gravitational effective field theory islands, low-spin dominance, and the four-graviton amplitude},'' \href{http://dx.doi.org/10.1088/1751-8121/ac0e51}{{\em J. Phys. A} {\bfseries 54} no.~34, (2021) 344002}, \href{http://arxiv.org/abs/2103.12728}{{\ttfamily arXiv:2103.12728 [hep-th]}}.

\bibitem{Li:2022rag}
X.~Li, K.~Mimasu, K.~Yamashita, C.~Yang, C.~Zhang, and S.-Y. Zhou, ``{Moments for positivity: using Drell-Yan data to test positivity bounds and reverse-engineer new physics},'' \href{http://dx.doi.org/10.1007/JHEP10(2022)107}{{\em JHEP} {\bfseries 10} (2022) 107}, \href{http://arxiv.org/abs/2204.13121}{{\ttfamily arXiv:2204.13121 [hep-ph]}}.

\bibitem{Caron-Huot:2022ugt}
S.~Caron-Huot, Y.-Z. Li, J.~Parra-Martinez, and D.~Simmons-Duffin, ``{Causality constraints on corrections to Einstein gravity},'' \href{http://dx.doi.org/10.1007/JHEP05(2023)122}{{\em JHEP} {\bfseries 05} (2023) 122}, \href{http://arxiv.org/abs/2201.06602}{{\ttfamily arXiv:2201.06602 [hep-th]}}.

\bibitem{Herrero-Valea:2020wxz}
M.~Herrero-Valea, R.~Santos-Garcia, and A.~Tokareva, ``{Massless positivity in graviton exchange},'' \href{http://dx.doi.org/10.1103/PhysRevD.104.085022}{{\em Phys. Rev. D} {\bfseries 104} no.~8, (2021) 085022}, \href{http://arxiv.org/abs/2011.11652}{{\ttfamily arXiv:2011.11652 [hep-th]}}.

\bibitem{EliasMiro:2022xaa}
J.~Elias~Miro, A.~Guerrieri, and M.~A. Gumus, ``{Bridging positivity and S-matrix bootstrap bounds},'' \href{http://dx.doi.org/10.1007/JHEP05(2023)001}{{\em JHEP} {\bfseries 05} (2023) 001}, \href{http://arxiv.org/abs/2210.01502}{{\ttfamily arXiv:2210.01502 [hep-th]}}.

\bibitem{Bellazzini:2021oaj}
B.~Bellazzini, M.~Riembau, and F.~Riva, ``{IR side of positivity bounds},'' \href{http://dx.doi.org/10.1103/PhysRevD.106.105008}{{\em Phys. Rev. D} {\bfseries 106} no.~10, (2022) 105008}, \href{http://arxiv.org/abs/2112.12561}{{\ttfamily arXiv:2112.12561 [hep-th]}}.

\bibitem{Sinha:2020win}
A.~Sinha and A.~Zahed, ``{Crossing Symmetric Dispersion Relations in Quantum Field Theories},'' \href{http://dx.doi.org/10.1103/PhysRevLett.126.181601}{{\em Phys. Rev. Lett.} {\bfseries 126} no.~18, (2021) 181601}, \href{http://arxiv.org/abs/2012.04877}{{\ttfamily arXiv:2012.04877 [hep-th]}}.

\bibitem{Trott:2020ebl}
T.~Trott, ``{Causality, unitarity and symmetry in effective field theory},'' \href{http://dx.doi.org/10.1007/JHEP07(2021)143}{{\em JHEP} {\bfseries 07} (2021) 143}, \href{http://arxiv.org/abs/2011.10058}{{\ttfamily arXiv:2011.10058 [hep-ph]}}.

\bibitem{Herrero-Valea:2022lfd}
M.~Herrero-Valea, A.~S. Koshelev, and A.~Tokareva, ``{UV graviton scattering and positivity bounds from IR dispersion relations},'' \href{http://dx.doi.org/10.1103/PhysRevD.106.105002}{{\em Phys. Rev. D} {\bfseries 106} no.~10, (2022) 105002}, \href{http://arxiv.org/abs/2205.13332}{{\ttfamily arXiv:2205.13332 [hep-th]}}.

\bibitem{Hong:2023zgm}
D.-Y. Hong, Z.-H. Wang, and S.-Y. Zhou, ``{Causality bounds on scalar-tensor EFTs},'' \href{http://dx.doi.org/10.1007/JHEP10(2023)135}{{\em JHEP} {\bfseries 10} (2023) 135}, \href{http://arxiv.org/abs/2304.01259}{{\ttfamily arXiv:2304.01259 [hep-th]}}.

\bibitem{Chiang:2022jep}
L.-Y. Chiang, Y.-t. Huang, W.~Li, L.~Rodina, and H.-C. Weng, ``{(Non)-projective bounds on gravitational EFT},'' \href{http://arxiv.org/abs/2201.07177}{{\ttfamily arXiv:2201.07177 [hep-th]}}.

\bibitem{Huang:2020nqy}
Y.-t. Huang, J.-Y. Liu, L.~Rodina, and Y.~Wang, ``{Carving out the Space of Open-String S-matrix},'' \href{http://dx.doi.org/10.1007/JHEP04(2021)195}{{\em JHEP} {\bfseries 04} (2021) 195}, \href{http://arxiv.org/abs/2008.02293}{{\ttfamily arXiv:2008.02293 [hep-th]}}.

\bibitem{Noumi:2021uuv}
T.~Noumi and J.~Tokuda, ``{Gravitational positivity bounds on scalar potentials},'' \href{http://dx.doi.org/10.1103/PhysRevD.104.066022}{{\em Phys. Rev. D} {\bfseries 104} no.~6, (2021) 066022}, \href{http://arxiv.org/abs/2105.01436}{{\ttfamily arXiv:2105.01436 [hep-th]}}.

\bibitem{Xu:2023lpq}
H.~Xu and S.-Y. Zhou, ``{Triple crossing positivity bounds, mass dependence and cosmological scalars: Horndeski theory and DHOST},'' \href{http://dx.doi.org/10.1088/1475-7516/2023/11/076}{{\em JCAP} {\bfseries 11} (2023) 076}, \href{http://arxiv.org/abs/2306.06639}{{\ttfamily arXiv:2306.06639 [hep-th]}}.

\bibitem{Chen:2023bhu}
Q.~Chen, K.~Mimasu, T.~A. Wu, G.-D. Zhang, and S.-Y. Zhou, ``{Capping the positivity cone: dimension-8 Higgs operators in the SMEFT},'' \href{http://dx.doi.org/10.1007/JHEP03(2024)180}{{\em JHEP} {\bfseries 03} (2024) 180}, \href{http://arxiv.org/abs/2309.15922}{{\ttfamily arXiv:2309.15922 [hep-ph]}}.

\bibitem{Noumi:2022wwf}
T.~Noumi and J.~Tokuda, ``{Finite energy sum rules for gravitational Regge amplitudes},'' \href{http://dx.doi.org/10.1007/JHEP06(2023)032}{{\em JHEP} {\bfseries 06} (2023) 032}, \href{http://arxiv.org/abs/2212.08001}{{\ttfamily arXiv:2212.08001 [hep-th]}}.

\bibitem{deRham:2022hpx}
C.~de~Rham, S.~Kundu, M.~Reece, A.~J. Tolley, and S.-Y. Zhou, ``{Snowmass White Paper: UV Constraints on IR Physics},'' in {\em {Snowmass 2021}}.
\newblock 3, 2022.
\newblock \href{http://arxiv.org/abs/2203.06805}{{\ttfamily arXiv:2203.06805 [hep-th]}}.

\bibitem{Hong:2024fbl}
D.-Y. Hong, Z.-H. Wang, and S.-Y. Zhou, ``{On Capped Higgs Positivity Cone},''
\newblock 4, 2024.
\newblock \href{http://arxiv.org/abs/2404.04479}{{\ttfamily arXiv:2404.04479 [hep-ph]}}.

\bibitem{Bern:2022yes}
Z.~Bern, E.~Herrmann, D.~Kosmopoulos, and R.~Roiban, ``{Effective Field Theory islands from perturbative and nonperturbative four-graviton amplitudes},'' \href{http://dx.doi.org/10.1007/JHEP01(2023)113}{{\em JHEP} {\bfseries 01} (2023) 113}, \href{http://arxiv.org/abs/2205.01655}{{\ttfamily arXiv:2205.01655 [hep-th]}}.

\bibitem{Ma:2023vgc}
T.~Ma, A.~Pomarol, and F.~Sciotti, ``{Bootstrapping the chiral anomaly at large N$_{c}$},'' \href{http://dx.doi.org/10.1007/JHEP11(2023)176}{{\em JHEP} {\bfseries 11} (2023) 176}, \href{http://arxiv.org/abs/2307.04729}{{\ttfamily arXiv:2307.04729 [hep-th]}}.

\bibitem{DeAngelis:2023bmd}
S.~De~Angelis and G.~Durieux, ``{EFT matching from analyticity and unitarity},'' \href{http://dx.doi.org/10.21468/SciPostPhys.16.3.071}{{\em SciPost Phys.} {\bfseries 16} (2024) 071}, \href{http://arxiv.org/abs/2308.00035}{{\ttfamily arXiv:2308.00035 [hep-ph]}}.

\bibitem{Acanfora:2023axz}
F.~Acanfora, A.~Guerrieri, K.~H\"aring, and D.~Karateev, ``{Bounds on scattering of neutral Goldstones},'' \href{http://dx.doi.org/10.1007/JHEP03(2024)028}{{\em JHEP} {\bfseries 03} (2024) 028}, \href{http://arxiv.org/abs/2310.06027}{{\ttfamily arXiv:2310.06027 [hep-th]}}.

\bibitem{Aoki:2023khq}
K.~Aoki, T.~Noumi, R.~Saito, S.~Sato, S.~Shirai, J.~Tokuda, and M.~Yamazaki, ``{Gravitational positivity for phenomenologists: Dark gauge boson in the swampland},'' \href{http://dx.doi.org/10.1103/PhysRevD.110.016002}{{\em Phys. Rev. D} {\bfseries 110} no.~1, (2024) 016002}, \href{http://arxiv.org/abs/2305.10058}{{\ttfamily arXiv:2305.10058 [hep-ph]}}.

\bibitem{Xu:2024iao}
H.~Xu, D.-Y. Hong, Z.-H. Wang, and S.-Y. Zhou, ``{Positivity bounds on parity-violating scalar-tensor EFTs},'' \href{http://dx.doi.org/10.1088/1475-7516/2025/01/102}{{\em JCAP} {\bfseries 01} (2025) 102}, \href{http://arxiv.org/abs/2410.09794}{{\ttfamily arXiv:2410.09794 [hep-th]}}.

\bibitem{EliasMiro:2023fqi}
J.~Elias~Miro, A.~L. Guerrieri, and M.~A. Gumus, ``{Extremal Higgs couplings},'' \href{http://dx.doi.org/10.1103/PhysRevD.110.016007}{{\em Phys. Rev. D} {\bfseries 110} no.~1, (2024) 016007}, \href{http://arxiv.org/abs/2311.09283}{{\ttfamily arXiv:2311.09283 [hep-ph]}}.

\bibitem{McPeak:2023wmq}
B.~McPeak, M.~Venuti, and A.~Vichi, ``{Adding subtractions: comparing the impact of different Regge behaviors},'' \href{http://arxiv.org/abs/2310.06888}{{\ttfamily arXiv:2310.06888 [hep-th]}}.

\bibitem{Riembau:2022yse}
M.~Riembau, ``{Full Unitarity and the Moments of Scattering Amplitudes},'' \href{http://arxiv.org/abs/2212.14056}{{\ttfamily arXiv:2212.14056 [hep-th]}}.

\bibitem{Caron-Huot:2024tsk}
S.~Caron-Huot and J.~Tokuda, ``{String loops and gravitational positivity bounds: imprint of light particles at high energies},'' \href{http://dx.doi.org/10.1007/JHEP11(2024)055}{{\em JHEP} {\bfseries 11} (2024) 055}, \href{http://arxiv.org/abs/2406.07606}{{\ttfamily arXiv:2406.07606 [hep-th]}}.

\bibitem{Caron-Huot:2024lbf}
S.~Caron-Huot and Y.-Z. Li, ``{Gravity and a universal cutoff for field theory},'' \href{http://arxiv.org/abs/2408.06440}{{\ttfamily arXiv:2408.06440 [hep-th]}}.

\bibitem{Wan:2024eto}
S.-L. Wan and S.-Y. Zhou, ``{Matrix moment approach to positivity bounds and UV reconstruction from IR},'' \href{http://arxiv.org/abs/2411.11964}{{\ttfamily arXiv:2411.11964 [hep-th]}}.

\bibitem{Buoninfante:2024ibt}
L.~Buoninfante, L.-Q. Shao, and A.~Tokareva, ``{Non-local positivity bounds: islands in Terra Incognita},'' \href{http://arxiv.org/abs/2412.08634}{{\ttfamily arXiv:2412.08634 [hep-th]}}.

\bibitem{Berman:2024owc}
J.~Berman and N.~Geiser, ``{Analytic bootstrap bounds on masses and spins in gravitational and non-gravitational scalar theories},'' \href{http://arxiv.org/abs/2412.17902}{{\ttfamily arXiv:2412.17902 [hep-th]}}.

\bibitem{deRham:2025vaq}
C.~de~Rham, A.~J. Tolley, Z.-H. Wang, and S.-Y. Zhou, ``{Primal S-matrix bootstrap with dispersion relations},'' \href{http://arxiv.org/abs/2506.22546}{{\ttfamily arXiv:2506.22546 [hep-th]}}.

\bibitem{Berman:2025owb}
J.~Berman, H.~Elvang, and C.~Figueiredo, ``{Splitting Regions and Shrinking Islands from Higher Point Constraints},'' \href{http://arxiv.org/abs/2506.22538}{{\ttfamily arXiv:2506.22538 [hep-th]}}.

\bibitem{Bellazzini:2020cot}
B.~Bellazzini, J.~Elias~Mir\'o, R.~Rattazzi, M.~Riembau, and F.~Riva, ``{Positive moments for scattering amplitudes},'' \href{http://dx.doi.org/10.1103/PhysRevD.104.036006}{{\em Phys. Rev. D} {\bfseries 104} no.~3, (2021) 036006}, \href{http://arxiv.org/abs/2011.00037}{{\ttfamily arXiv:2011.00037 [hep-th]}}.

\bibitem{Beadle:2024hqg}
C.~Beadle, G.~Isabella, D.~Perrone, S.~Ricossa, F.~Riva, and F.~Serra, ``{Non-Forward UV/IR Relations},'' \href{http://arxiv.org/abs/2407.02346}{{\ttfamily arXiv:2407.02346 [hep-th]}}.

\bibitem{Ye:2024rzr}
Y.~Ye, B.~He, and J.~Gu, ``{Positivity bounds in scalar Effective Field Theories at one-loop level},'' \href{http://dx.doi.org/10.1007/JHEP12(2024)046}{{\em JHEP} {\bfseries 12} (2024) 046}, \href{http://arxiv.org/abs/2408.10318}{{\ttfamily arXiv:2408.10318 [hep-ph]}}.

\bibitem{Bertucci:2024qzt}
F.~Bertucci, J.~Henriksson, B.~McPeak, S.~Ricossa, F.~Riva, and A.~Vichi, ``{Positivity bounds on massive vectors},'' \href{http://dx.doi.org/10.1007/JHEP12(2024)051}{{\em JHEP} {\bfseries 12} (2024) 051}, \href{http://arxiv.org/abs/2402.13327}{{\ttfamily arXiv:2402.13327 [hep-th]}}.

\bibitem{Chang:2025cxc}
C.-H. Chang and J.~Parra-Martinez, ``{Graviton loops and negativity},'' \href{http://arxiv.org/abs/2501.17949}{{\ttfamily arXiv:2501.17949 [hep-th]}}.

\bibitem{Beadle:2025cdx}
C.~Beadle, G.~Isabella, D.~Perrone, S.~Ricossa, F.~Riva, and F.~Serra, ``{The EFT Bootstrap at Finite $M_{PL}$},'' \href{http://arxiv.org/abs/2501.18465}{{\ttfamily arXiv:2501.18465 [hep-th]}}.

\bibitem{Pasiecznik:2025eqc}
C.~Pasiecznik, ``{Bootstrapping Gravity with Crossing Symmetric Dispersion Relations},'' \href{http://arxiv.org/abs/2506.09884}{{\ttfamily arXiv:2506.09884 [hep-th]}}.

\bibitem{Ye:2025zhs}
Y.~Ye, X.~Cao, Y.-H. Wu, and J.~Gu, ``{Positivity bounds in scalar-QED EFT at one-loop level},'' \href{http://arxiv.org/abs/2507.06302}{{\ttfamily arXiv:2507.06302 [hep-ph]}}.

\bibitem{Andriolo:2020lul}
S.~Andriolo, T.-C. Huang, T.~Noumi, H.~Ooguri, and G.~Shiu, ``{Duality and axionic weak gravity},'' \href{http://dx.doi.org/10.1103/PhysRevD.102.046008}{{\em Phys. Rev. D} {\bfseries 102} no.~4, (2020) 046008}, \href{http://arxiv.org/abs/2004.13721}{{\ttfamily arXiv:2004.13721 [hep-th]}}.

\bibitem{Henriksson:2022oeu}
J.~Henriksson, B.~McPeak, F.~Russo, and A.~Vichi, ``{Bounding violations of the weak gravity conjecture},'' \href{http://dx.doi.org/10.1007/JHEP08(2022)184}{{\em JHEP} {\bfseries 08} (2022) 184}, \href{http://arxiv.org/abs/2203.08164}{{\ttfamily arXiv:2203.08164 [hep-th]}}.

\bibitem{Aoki:2021ckh}
K.~Aoki, T.~Q. Loc, T.~Noumi, and J.~Tokuda, ``{Is the Standard Model in the Swampland? Consistency Requirements from Gravitational Scattering},'' \href{http://dx.doi.org/10.1103/PhysRevLett.127.091602}{{\em Phys. Rev. Lett.} {\bfseries 127} no.~9, (2021) 091602}, \href{http://arxiv.org/abs/2104.09682}{{\ttfamily arXiv:2104.09682 [hep-th]}}.

\bibitem{Alberte:2021dnj}
L.~Alberte, C.~de~Rham, S.~Jaitly, and A.~J. Tolley, ``{Reverse Bootstrapping: IR Lessons for UV Physics},'' \href{http://dx.doi.org/10.1103/PhysRevLett.128.051602}{{\em Phys. Rev. Lett.} {\bfseries 128} no.~5, (2022) 051602}, \href{http://arxiv.org/abs/2111.09226}{{\ttfamily arXiv:2111.09226 [hep-th]}}.

\bibitem{deRham:2022gfe}
C.~de~Rham, S.~Jaitly, and A.~J. Tolley, ``{Constraints on Regge behavior from IR physics},'' \href{http://dx.doi.org/10.1103/PhysRevD.108.046011}{{\em Phys. Rev. D} {\bfseries 108} no.~4, (2023) 046011}, \href{http://arxiv.org/abs/2212.04975}{{\ttfamily arXiv:2212.04975 [hep-th]}}.

\bibitem{Haring:2022cyf}
K.~H\"aring and A.~Zhiboedov, ``{Gravitational Regge bounds},'' \href{http://dx.doi.org/10.21468/SciPostPhys.16.1.034}{{\em SciPost Phys.} {\bfseries 16} no.~1, (2024) 034}, \href{http://arxiv.org/abs/2202.08280}{{\ttfamily arXiv:2202.08280 [hep-th]}}.

\bibitem{Haring:2024wyz}
K.~H\"aring and A.~Zhiboedov, ``{What is the graviton pole made of?},'' \href{http://arxiv.org/abs/2410.21499}{{\ttfamily arXiv:2410.21499 [hep-th]}}.

\bibitem{Froissart:1961ux}
M.~Froissart, ``{Asymptotic behavior and subtractions in the Mandelstam representation},'' \href{http://dx.doi.org/10.1103/PhysRev.123.1053}{{\em Phys. Rev.} {\bfseries 123} (1961) 1053--1057}.

\bibitem{Martin:1962rt}
A.~Martin, ``{Unitarity and high-energy behavior of scattering amplitudes},'' \href{http://dx.doi.org/10.1103/PhysRev.129.1432}{{\em Phys. Rev.} {\bfseries 129} (1963) 1432--1436}.

\bibitem{Jin:1964zz}
Y.~S. Jin and A.~Martin, ``{Connection Between the Asymptotic Behavior and the Sign of the Discontinuity in One-Dimensional Dispersion Relations},'' \href{http://dx.doi.org/10.1103/PhysRev.135.B1369}{{\em Phys. Rev.} {\bfseries 135} (1964) B1369--B1374}.

\bibitem{Arkani-Hamed:2020blm}
N.~Arkani-Hamed, T.-C. Huang, and Y.-t. Huang, ``{The EFT-Hedron},'' \href{http://dx.doi.org/10.1007/JHEP05(2021)259}{{\em JHEP} {\bfseries 05} (2021) 259}, \href{http://arxiv.org/abs/2012.15849}{{\ttfamily arXiv:2012.15849 [hep-th]}}.

\bibitem{Chiang:2021ziz}
L.-Y. Chiang, Y.-t. Huang, W.~Li, L.~Rodina, and H.-C. Weng, ``{Into the EFThedron and UV constraints from IR consistency},'' \href{http://dx.doi.org/10.1007/JHEP03(2022)063}{{\em JHEP} {\bfseries 03} (2022) 063}, \href{http://arxiv.org/abs/2105.02862}{{\ttfamily arXiv:2105.02862 [hep-th]}}.

\bibitem{Chiang:2022ltp}
L.-Y. Chiang, Y.-t. Huang, L.~Rodina, and H.-C. Weng, ``{De-projecting the EFThedron},'' \href{http://dx.doi.org/10.1007/JHEP05(2024)102}{{\em JHEP} {\bfseries 05} (2024) 102}, \href{http://arxiv.org/abs/2204.07140}{{\ttfamily arXiv:2204.07140 [hep-th]}}.

\end{thebibliography}\endgroup


\end{document}